\documentclass[preprint2,twocolumn,times,tighten]{aastex63}

\usepackage{graphicx,times}
\usepackage{subfigure}
\newcommand{\be}{\begin{equation}}
\usepackage{threeparttable}
\usepackage{booktabs}
\newcommand{\ee}{\end{equation}}
\newcommand{\bea}{\begin{eqnarray}}
\newcommand{\eea}{\end{eqnarray}}
\newcommand*{\xu}{\color{black}}

\usepackage{enumerate}
\usepackage{amsmath}
\usepackage{cases}
\usepackage{longtable}
\usepackage{hyperref}
\usepackage{epstopdf}
\usepackage{amsmath,bm}
\usepackage{amssymb}
\usepackage{natbib}
\usepackage{morefloats}
\usepackage{multirow}
\usepackage{array}
\usepackage{verbatim}
\usepackage{lineno}

\shorttitle{}
\shortauthors{Xu et al.}

\begin{document}

\title{Statistical measurements of dispersion measure fluctuations of FRBs}

\email{Hubble Fellow, sxu@ias.edu; weinberg.21@osu.edu; bing.zhang@unlv.edu}

\author[0000-0002-5771-2055]{Siyao Xu}
\affiliation{Institute for Advanced Study, 1 Einstein Drive, Princeton, NJ 08540, USA}

\author[0000-0001-7775-7261]{David H.~Weinberg}
\affiliation{Department of Astronomy and Center for Cosmology
and AstroParticle Physics, The Ohio State University,
Columbus, OH 43210, USA}
\affiliation{Institute for Advanced Study, 1 Einstein Drive, Princeton, NJ 08540, USA}

\author[0000-0002-9725-2524]{Bing Zhang}
\affiliation{Department of Physics and Astronomy, University of Nevada Las Vegas, NV 89154, USA}

\begin{abstract}
Extragalactic 
Fast Radio Bursts (FRBs) have large dispersion measures (DMs) and are unique probes of intergalactic electron density fluctuations. 
By using the recently released First CHIME/FRB Catalog, we reexamined the structure function (SF) of DM fluctuations. 
It shows a large DM fluctuation similar to that previously reported in
\citet{XuZfrb20}, 
but no clear correlation hinting towards large scale turbulence is reproduced with this larger sample.
To suppress the distortion effect from FRB distances and their host DMs, 
we focus on a subset of CHIME catalog
with DM $<500$ pc cm$^{-3}$.
A trend of 
non-constant SF and non-zero correlation function (CF) at angular separations $\theta$ less than $10^\circ$ is seen, but with large statistical uncertainties. 
The difference found between SF and 
{\xu that derived from CF}
at $\theta \lesssim 10^\circ$
can be ascribed to the large 
statistical uncertainties 
or the density inhomogeneities on scales on the order of $100$ Mpc.
The possible correlation of electron density fluctuations and inhomogeneities of density distribution 
should be tested when 
{\xu several thousands of} FRBs are available. 

\end{abstract}


\section{Introduction}

Extragalactic 
Fast Radio Bursts (FRBs) 
\citep{Lor07,Tho13,Pat16}
have their dispersion measures (DMs) 
greatly exceeding those of Galactic pulsars in the high Galactic latitude region 
\citep{Cord19}. 
Thus they provide unique 
probes of the intergalactic electron 
density fluctuations 
\citep{Macq13,XZ16,Rav16,Zhu18}.
The cosmological applications of FRBs on, e.g., constraining  
the baryon content of the Universe 
\citep{Keane16, Macqnat20}
and cosmological parameters
\citep{Deng14,Gao14,Zhou14,Walters18,Kumar19} via DM-redshift relation, 
probing reionization history of the universe 
\citep{Deng14,Zheng14,Caleb19,Beniamini21},
tracing the large-scale structure of the Universe
\citep{Masui15,Shiras17,Reisch21,rafi21},
have been extensively studied in the literature.

It is not straightforward to extract the intergalactic electron density fluctuations from the DM fluctuations of FRBs due to the unknown distances and host DMs. 
First, DMs are projected electron densities. 
Their fluctuations are affected and can even be dominated by dispersion of distances, depending on the line-of-sight thickness of the sample
\citep{LP16,XuZfrb20,XZpul20,Eina20}.
Second, the host DMs have a large dispersion 
\citep{Yang17}
{\xu and can be exceptionally large 
(up to $\sim 900$ pc cm$^{-3}$)
for some FRBs 
\citep{Niu21,rafi21}.}
In addition, with a limited sample size, 
the statistical measurements of DM fluctuations especially at small angular separations ($\theta < 10^\circ$) 
have large uncertainties, 
which makes the interpretation of the results 
more challenging
\citep{XuZfrb20}
(hereafter XZ20).

Theoretical models on DM fluctuations of FRBs 
based on 
cosmological hydrodynamic simulations
(e.g., \citealt{Takaha21})
provide insight into their statistical properties.  
Direct statistical 
measurements with observational data 
are important for testing the theoretical predictions and the underlying cosmological models. 
XZ20 made the first attempt 
and found a power-law structure function (SF) of DM fluctuations up to 
$100$ Mpc 
by using $112$ FRBs from 
FRBCAT
\citep{Pat16}
\footnote{http://www.frbcat.org},
but the {\xu statistical} uncertainties are very large due to the small sample size. 
The recently released 
First CHIME/FRB Catalog
\citep{chim21}
\footnote{https://www.chime-frb.ca/catalog}
with 535 FRBs provides us an opportunity to reexamine the SF 
with a larger FRB sample. 

The galaxy two-point 
correlation function 
\citep{Limber53,Peeb80,LanSza93}
is a powerful statistical tool for studying 
clustering of galaxies, clusters of galaxies, superclusters,
and testing cosmological models 
(e.g., \citealt{Klyp83,Bah86,Saun91,Zehavi02,Hawk03,Wein04,Wein13}).
Different from the galaxy correlation statistics that relates galaxies to the underlying mass distribution, the SF of DM fluctuations 
directly measures the statistical properties of the projected intergalactic electron density field.

The SF in general has a higher accuracy than the correlation function (CF) 
in the presence of noise and large-scale inhomogeneities
\citep{monin1965}. 
For a similar degree of accuracy, the SF requires a much smaller data size than the CF
\citep{Schul81}, which is an advantage for studying DMs of FRBs with a still limited sample size. 
The SF measurements have been applied to various observables related to velocities, densities, and magnetic fields in the multi-phase astrophysical media
(e.g., \citealt{XuZ16,Xu20,Li20}),
as well as velocities of {\xu young }stars 
\citep{Ha21}.
The applications of the SF to  large-scale intergalactic density fluctuations and studies on its  
cosmological implications 
are rare. 
In this work, 
we will further explore the SF of DM fluctuations by using 
the CHIME/FRB Catalog. 
{\xu We will examine 
whether the non-flat SF found from FRBCAT, which is indicative of the correlation of electron density fluctuations, can be still seen with a larger FRB sample. 
We will further 
compare the SF with the CF measurement,
to investigate whether they contain equivalent statistical information
as expected for homogeneous large-scale
intergalactic density distribution.} 
In Section 2, we first compare the SFs of DM fluctuations measured by 
XZ20 and measured with the new
CHIME Catalog. 
We then focus on the SF and CF analysis by using a subset of CHIME FRBs with relatively small DMs. 
The summary of our main results are given in Section 3.

\section{SF and CF of DM fluctuations of FRBs}

\subsection{Comparison between FRBCAT and CHIME samples}

\begin{figure}[htbp]
\centering   
   \includegraphics[width=8.5cm]{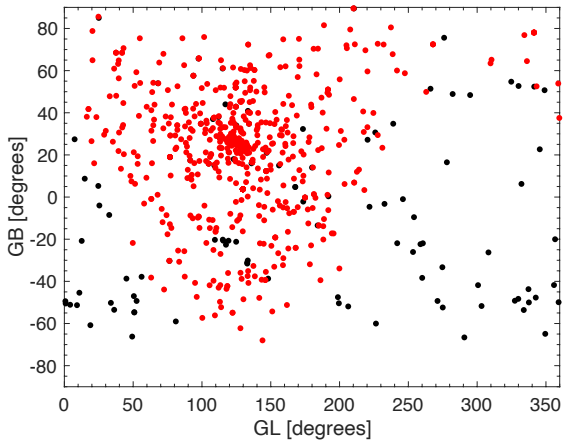}
\caption{$535$
FRBs from CHIME catalog in Galactic coordinates
(red)
overlaid on the $112$ FRBs 
from FRBCAT used in XZ20
(black).}
\label{fig:map}
\end{figure}

Fig. \ref{fig:map} shows the sky distribution of the FRBs from the CHIME Catalog and from FRBCAT. Some FRBs have their sky locations overlapped. 
The CHIME sample has a smaller sky coverage, but
contains more FRB pairs with smaller angular separations.

The SF of FRB DMs is defined as 
\begin{equation}
     D (\theta) = \langle (\text{DM} (\bm{X_1}) - \text{DM} (\bm{X_2}))^2 \rangle,
\end{equation}
where $\bm{X_1}$ and $\bm{X_2}$ are the sky locations of a pair of FRBs with the angular separation $\theta$, 
and $\langle...\rangle$ denotes the average over all pairs with the same $\theta$. 
The CF of DM fluctuations is
\begin{equation}
    \xi(\theta) = \langle \delta \text{DM}(\bm{X_1}) \delta \text{DM}(\bm{X_2})\rangle,
\end{equation}
where $\delta \text{DM} = \text{DM}-\overline{\text{DM}}$ is the DM fluctuation, and $\overline{\text{DM}}$ is the mean DM. 
We note that the SF of DMs and the SF of DM fluctuations are the same. 
For a stationary or a spatially homogeneous random process, the SF and CF contain equivalent information, and they 
are related by 
\begin{equation}\label{eq:sfcf}
    D(\theta) = \text{const} - 2 \xi(\theta). 
\end{equation}
In a more general case with low-frequency noise or large-scale fluctuations, 
the statistical characteristics of SF and CF are not mutually interchangeable. 
It was found in statistical mechanics that the SF is less distorted than the CF by 
large-scale inhomogeneities \citep{monin1965}.
To reach a similar accuracy, the CF requires a much larger (by one to two orders of magnitude)
data set than the SF
\citep{Schul81}.

\begin{figure*}[htbp]
\centering   
\subfigure[]{
\includegraphics[width=8cm]{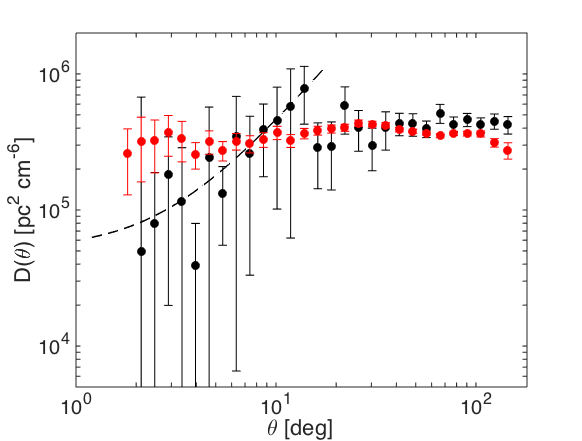}
   \label{fig: com1}}
\subfigure[]{
\includegraphics[width=8cm]{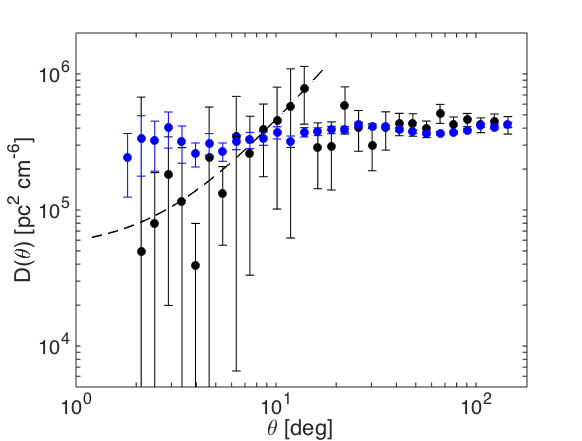}
   \label{fig: com2}}
\caption{
(a) $D(\theta)$ measured by XZ20
with $112$ FRBs 
from FRBCAT
(black)
and that measured with $491$
FRBs from CHIME
(red). 
Error bars indicate $95\%$ confidence intervals. 
The dashed line is the fit to the data points at small $\theta$ for the $112$ FRBs 
(XZ20).
(b) $D(\theta)$ measured with 
total $603$ FRBs from combined FRBCAT and CHIME samples (blue).}
\label{fig:oldnew}
\end{figure*}

Fig. \ref{fig: com1} presents
$D(\theta)$ measured by 
XZ20 
with $112$ FRBs from FRBCAT, 
in comparison with that measured with $491$ FRBs (in this work) with distinct coordinates
from CHIME. 
The uncertainty of the latter is much smaller due to the significantly larger sample size. 
We note that the error bars show $95\%$
confidence intervals, which are calculated at each $\theta$ using the Student's t-distribution when there are $<30$ FRB pairs and the normal distribution for more FRB pairs. 
The power-law trend indicated by the dashed line at small $\theta$ is not seen for the CHIME sample. 
We further combine the two catalogs,
and the corresponding $D(\theta)$ of total $603$ FRBs\footnote{Some FRBs in the two catalogs have overlapping sky positions.} 
is displayed in Fig. \ref{fig: com2}. 
We see that $D(\theta)$ is dominated by CHIME sample,
so the inclusion of FRBCAT FRBs does not significantly change the result, 
except for the very large-$\theta$ end. 

We believe that the different $D(\theta)$ measured with FRBCAT
and CHIME catalog at small $\theta$
is mainly caused by the different sample sizes. 
The latter measurements are within the uncertainty range of the former.
The effect of sky coverage is unclear and thus cannot be completely excluded.

\begin{figure*}[htbp]
\centering   
\subfigure[]{
\includegraphics[width=8cm]{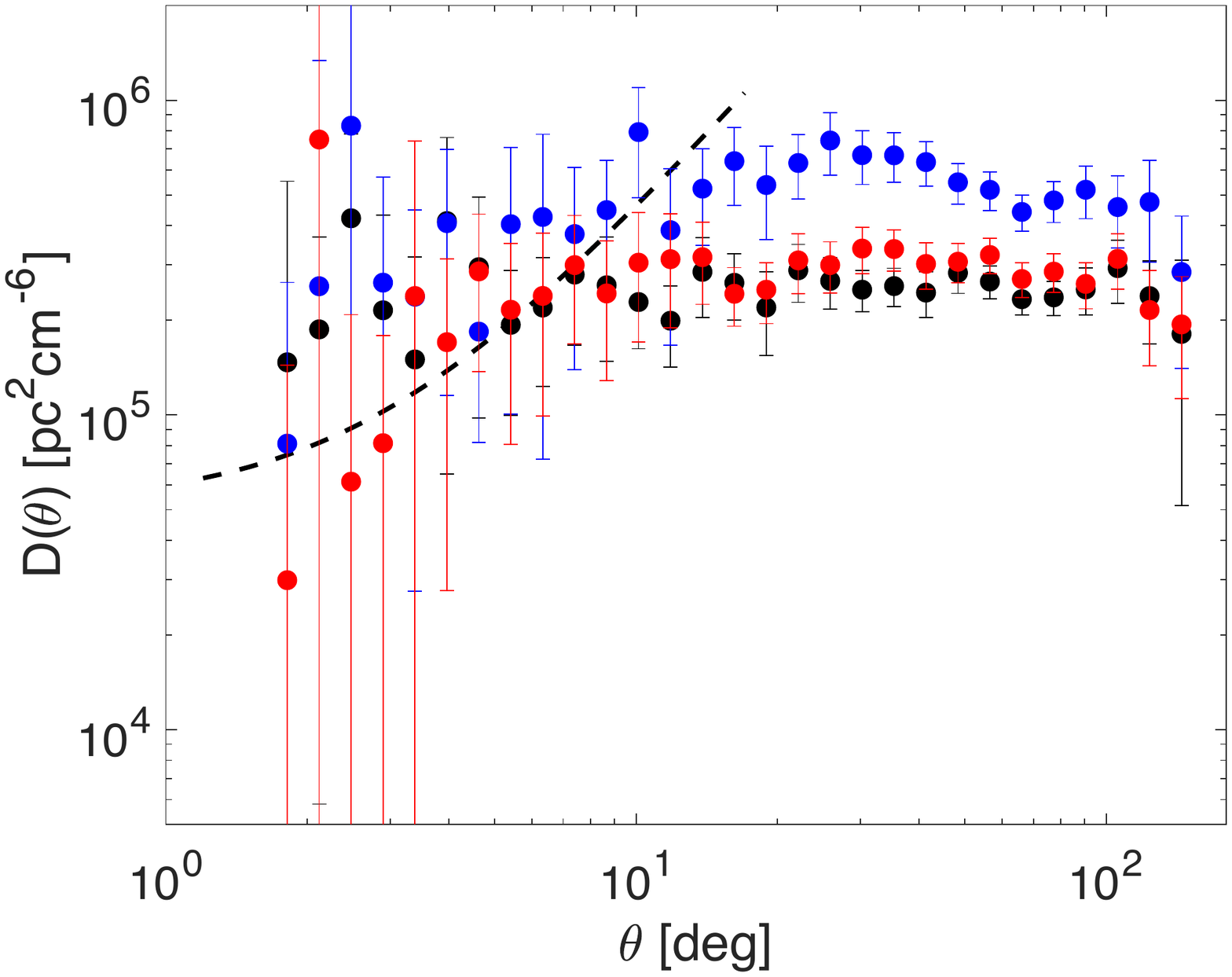}\label{fig: ran1}}
\subfigure[]{
\includegraphics[width=8cm]{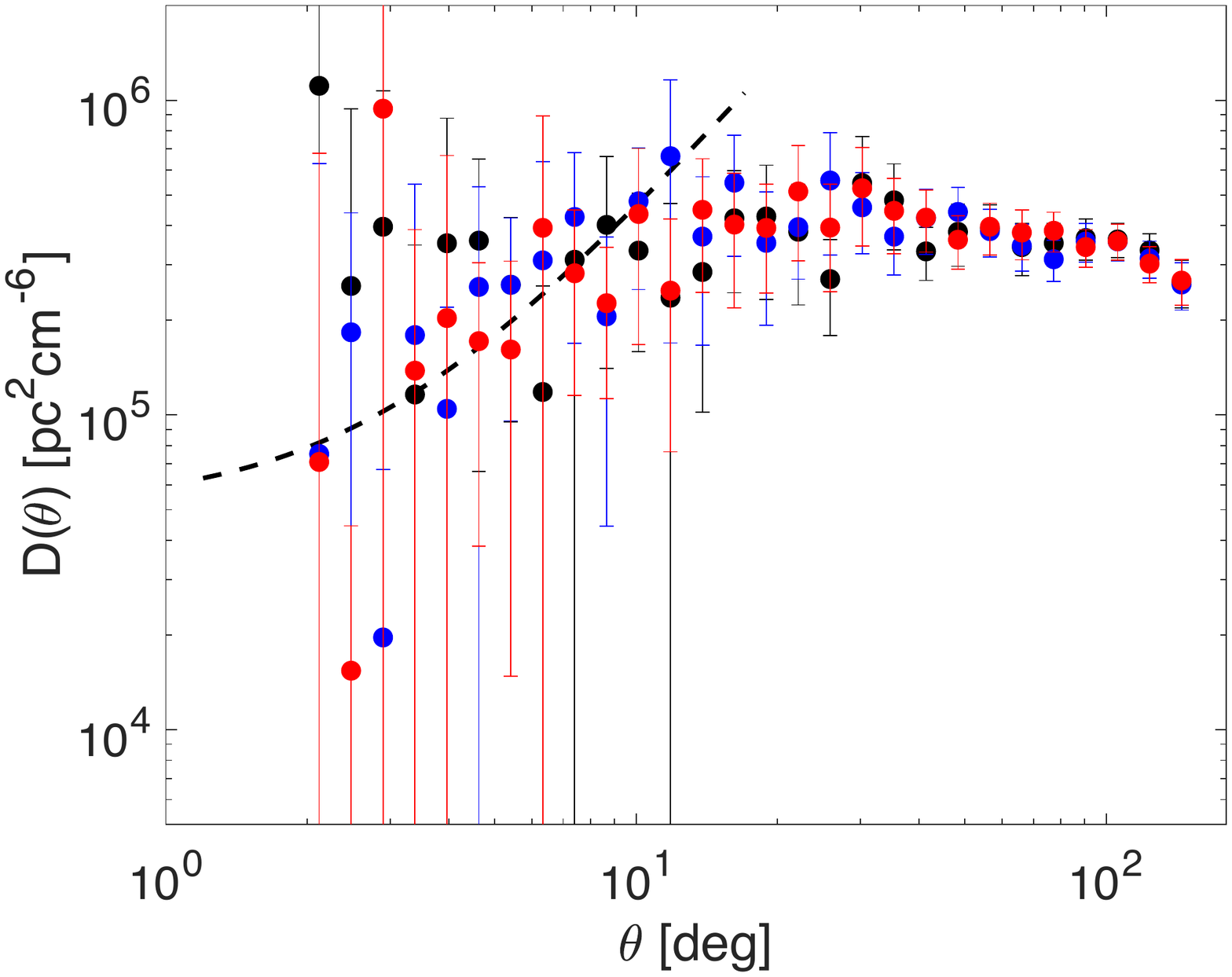}\label{fig: ran2}}
\caption{{\xu (a) $D(\theta)$ measured from a subsample of randomly selected $112$ FRBs
from CHIME.
(b) $D(\theta)$ measured from CHIME, but with the 
same number of FRB pairs in each $\theta$ bin as that for $D(\theta)$ measured from FRBCAT (see Fig. \ref{fig:oldnew}).
Error bars indicate $95\%$ confidence intervals.
Three colors represent three different
realizations.
The dashed line corresponds to the fit to $D(\theta)$ at small $\theta$
measured with FRBCAT (see Fig. \ref{fig:oldnew}). }
}
\label{fig:ranchime}
\end{figure*}

{\xu To further examine the effect of sample size on $D(\theta)$, we randomly select $112$ FRBs, i.e., the same number of FRBs as in FRBCAT, from the total CHIME sample as a subsample. 
As an example, 
$D(\theta)$ measured from three subsamples is shown in Fig. \ref{fig: ran1}. We see that
different subsamples can lead to different $D(\theta)$ at both small and large $\theta$. As an alternative test, 
we randomly select the same number of FRB pairs from the total CHIME sample in each $\theta$ bin as that for FRBCAT sample. 
$D(\theta)$ corresponding to three  
realizations  
is shown in Fig. \ref{fig: ran2} as an example. 
In both tests, we find
large statistical uncertainties of $D(\theta)$
at small $\theta$ because of the small sample size.
Occasionally, a power-law trend can be seen, e.g., blue circles in Fig. \ref{fig: ran2}.
It suggests that the power-law trend 
of $D(\theta)$ seen at small $\theta$ with a small sample of FRBs from FRBCAT is likely 
a coincidence.
}

\subsection{Statistics with a subset of CHIME sample}

\begin{figure}[htbp]
\centering   
   \includegraphics[width=8.5cm]{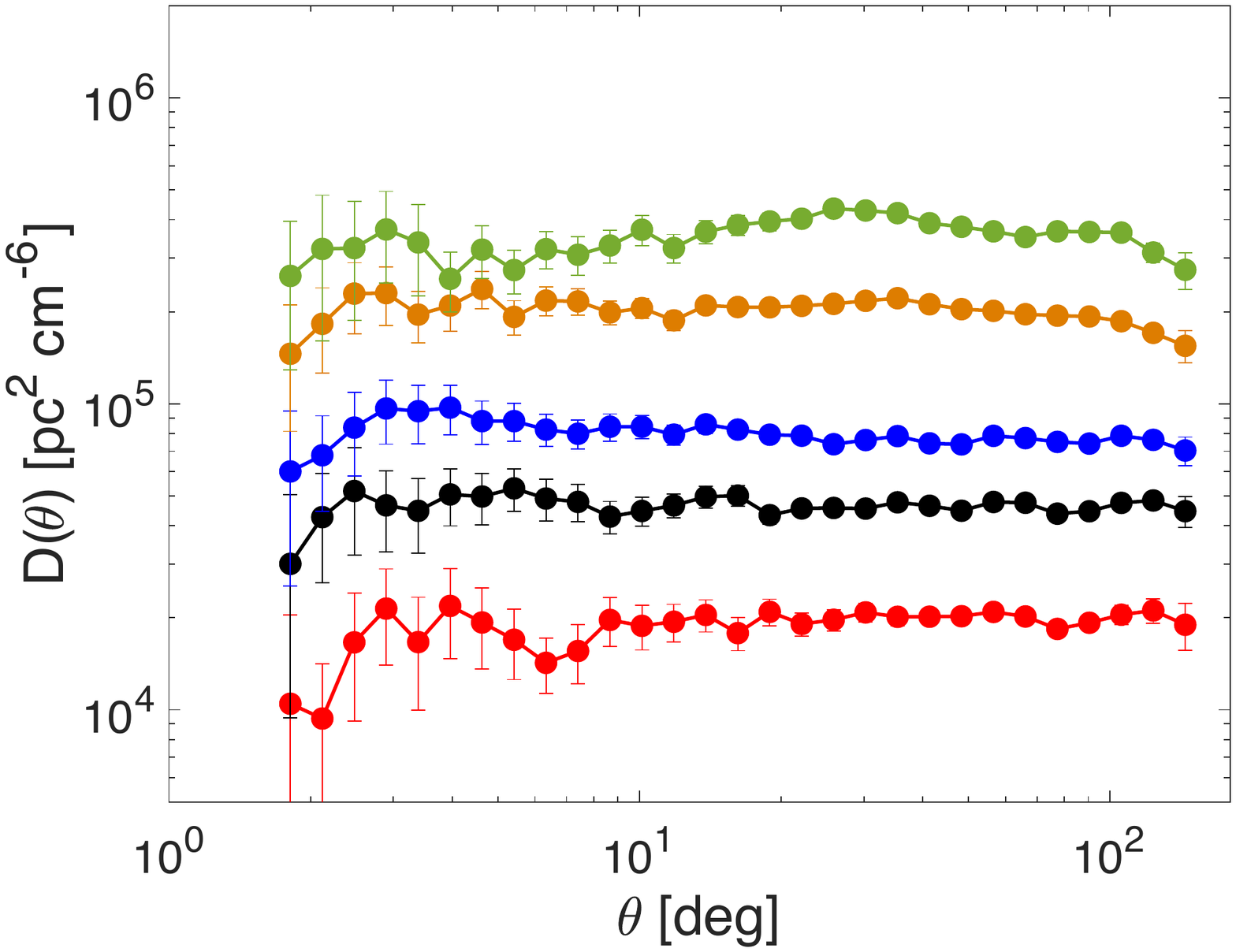}
\caption{$D(\theta)$ measured 
using a subset 
of the CHIME catalog, 
with 
$210$ FRBs with
$\text{DM}<500$ pc cm$^{-3}$ (red),
$322$ FRBs with
$\text{DM}<700$ pc cm$^{-3}$ (black),
$385$ FRBs with
$\text{DM}<900$ pc cm$^{-3}$ (blue),
$468$ FRBs with
$\text{DM}<1500$ pc cm$^{-3}$ (orange),
and $491$ FRBs without DM cut
(green).
{\xu Error bars indicate $95\%$ confidence intervals.}}
\label{fig: diffdmcut}
\end{figure}

Both
$D(\theta)$ and $\xi(\theta)$ of 
DM fluctuations are 2D projected statistical measurements of electron density fluctuations. 
The larger the dispersion of the distances to FRBs, the greater difference between the 2D and 3D statistics is expected. 
Depending on the thickness of the 2D sample, the correlation information can be partly or completely lost
\citep{Eina20}.
In addition, 
the host DMs of FRBs can largely distort the measured 
$D(\theta)$ and $\xi(\theta)$.
In particular, some FRBs can have very large host DMs ($\sim 400$ pc cm$^{-3}$), as suggested by 
\citet{rafi21}.
{\xu A repeating FRB with the host DM 
$\approx 902$ pc cm$^{-3}$
is recently found by 
\citet{Niu21}.}
As an attempt to constrain the thickness of the 2D sample and exclude FRBs with exceptionally large host DMs, 
we explore the possibility of using a subset of FRB sample 
by applying a DM cut for the statistical analysis. 
Fig. \ref{fig: diffdmcut} shows different $D(\theta)$ corresponding to different DM cuts. 
{\xu Naturally, the DM fluctuation of 
a subset of FRBs becomes smaller at a smaller DM cut value. 
}

\begin{figure}[htbp]
\centering   
   \includegraphics[width=8.5cm]{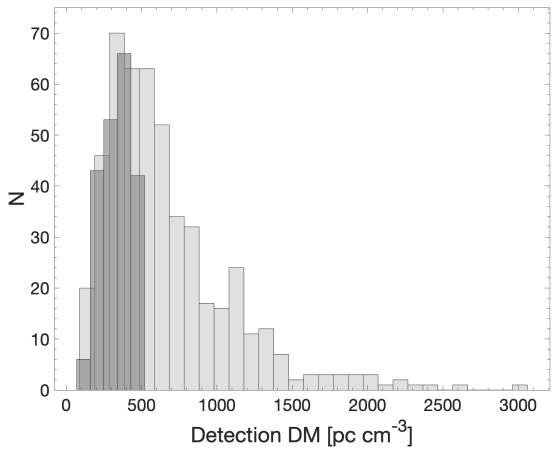}
\caption{
DM distribution for the entire CHIME catalog (light gray)
and for the subset 
with $\text{DM}<500$ pc cm$^{-3}$ (dark gray). }
\label{fig: pdf}
\end{figure}

There is a trade-off between 
the thickness of the 2D sample 
and the sample size. 
A higher cut value leads to a thicker 2D sample, and 
a lower one brings larger 
statistical uncertainties. 
We apply here a tentative DM cut at $500$ pc cm$^{-3}$
(see Fig. \ref{fig: pdf} for the corresponding DM distribution).
According to the DM-redshift relation
\citep{Deng14,Zha18},
\begin{equation}
     \text{DM}_\text{IGM} \approx 807~ \text{pc cm}^{-3} \int_0^z \frac{(1+z) dz}{ [\Omega_m (1+z)^3 + \Omega_\Lambda]^\frac{1}{2}},
\end{equation}
where $\Omega_m = 0.3089\pm 0.0062$ 
and $\Omega_\Lambda = 0.6911\pm0.0062$ are the matter density parameter and dark energy density parameter
\citep{Pla16},
we approximately sample the nearby density structures 
within redshift $z\approx0.57$ ($\approx 2\times10^3$ Mpc
as the LOS comoving distance)
by assuming the DM is dominated by its intergalactic component. 
The optimized selection cut
should be determined based on the accurate modeling of distances and host DMs of FRBs, 
as well as the sample size.

\begin{figure*}[htbp]
\centering   
\subfigure[Total detection DM]{
   \includegraphics[width=8.5cm]{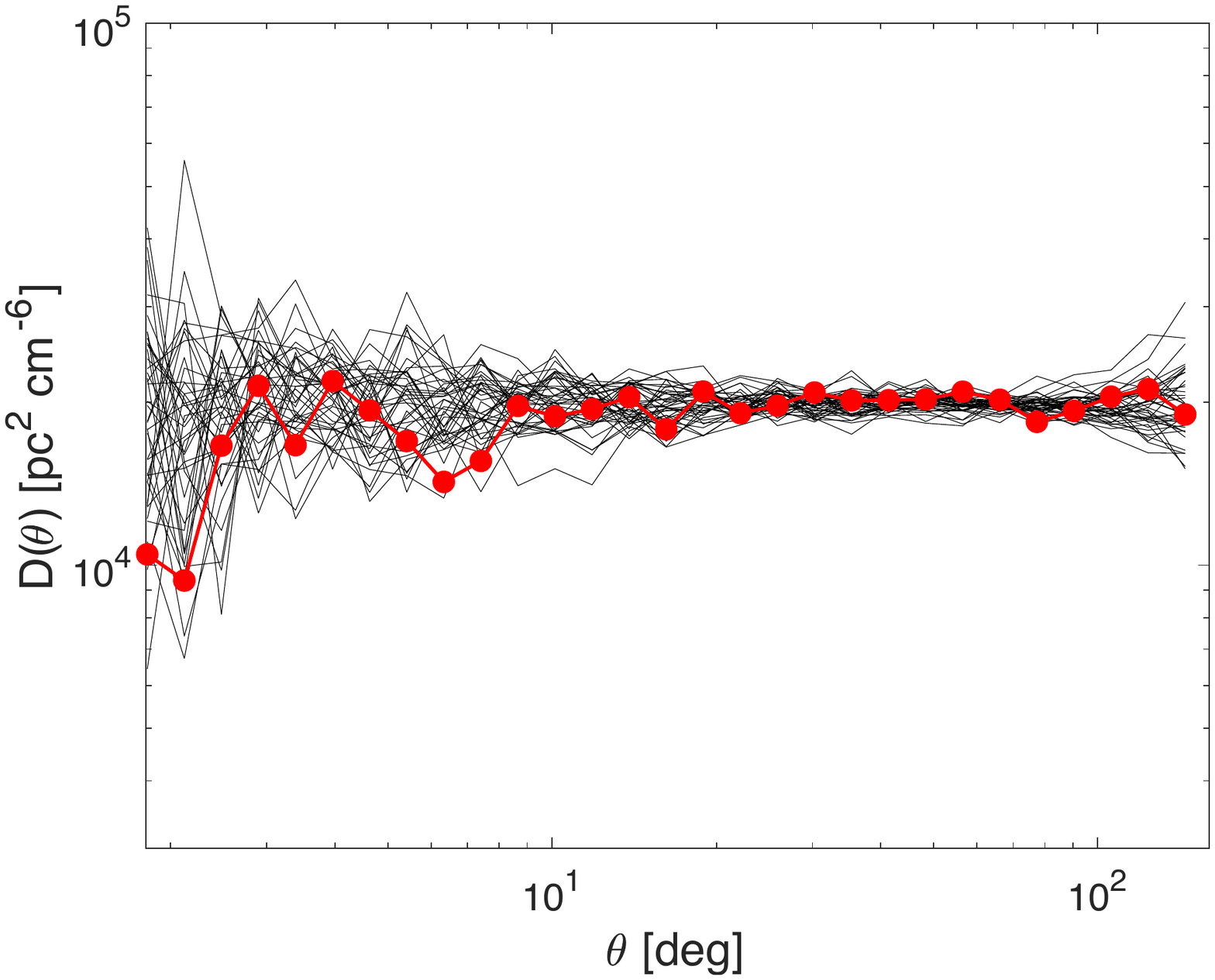}\label{fig: sftot}}
\subfigure[DM (NE2001)]{
   \includegraphics[width=8.5cm]{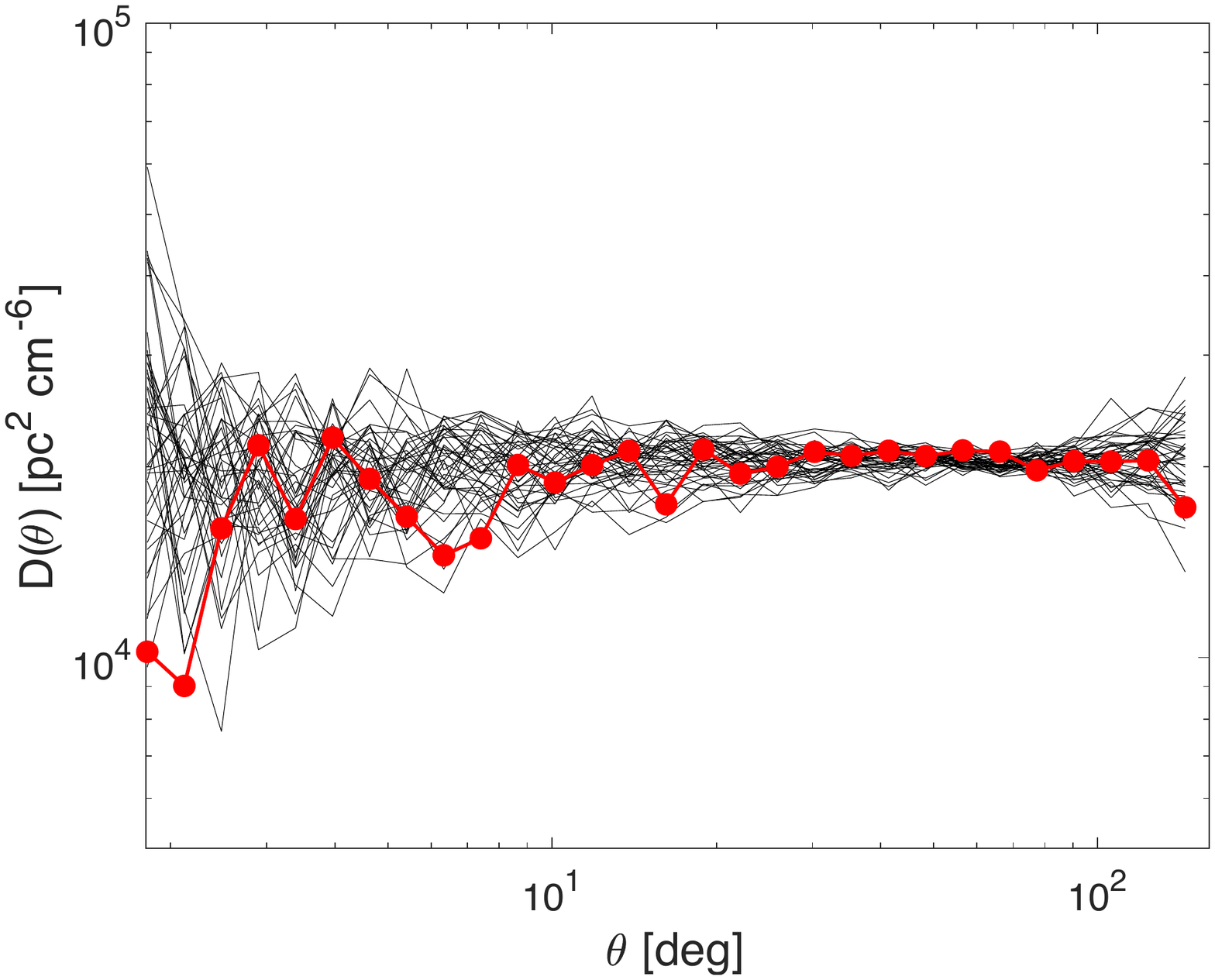}\label{fig: sfne}}
\subfigure[DM (YMW16)]{
   \includegraphics[width=8.5cm]{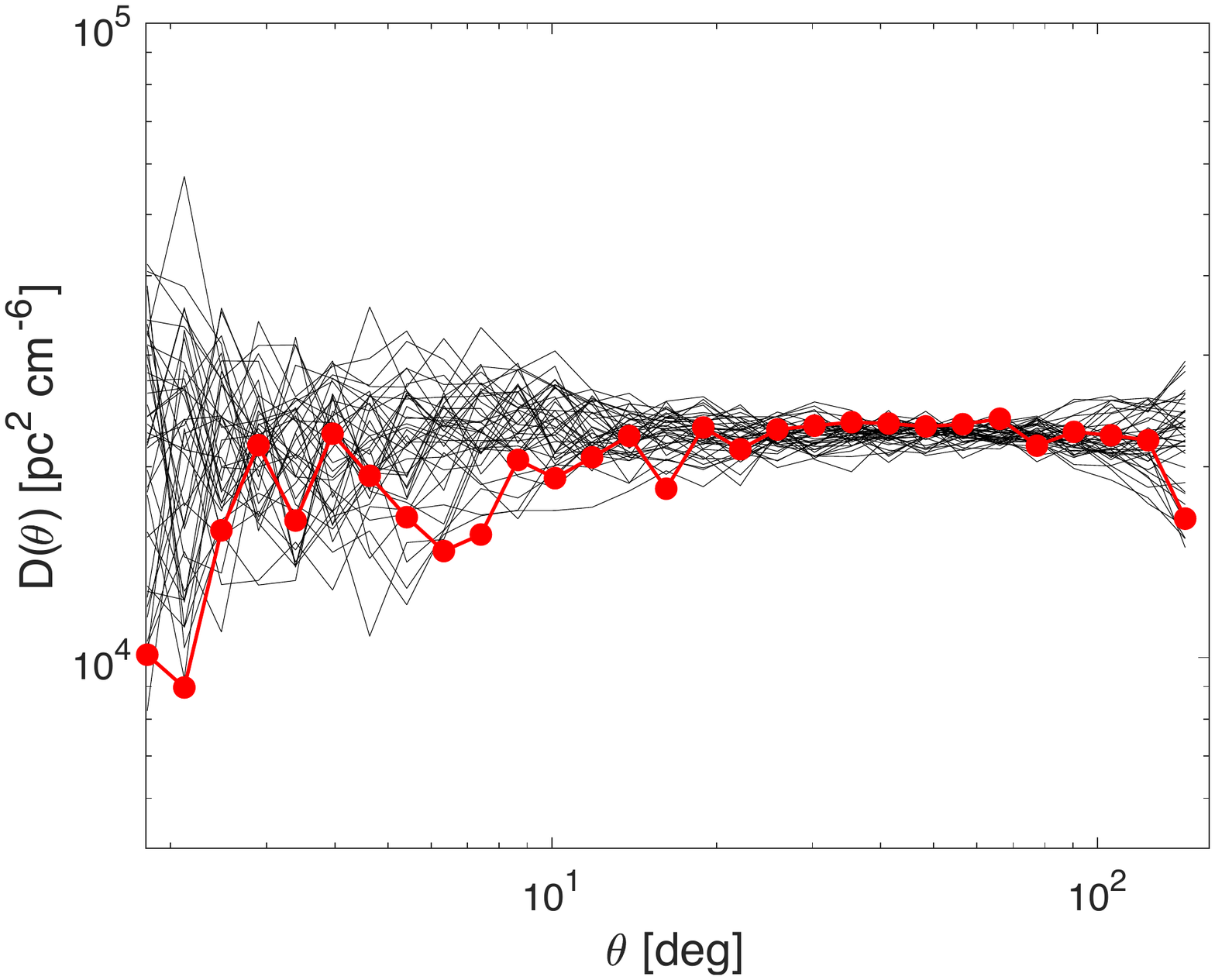}\label{fig: sfymw}}
\caption{ 
$D(\theta)$ vs. $\theta$ measured with
$210$ FRBs with total DM
$<500$ pc cm$^{-3}$,
corresponding to total DM in (a), 
DM (NE2001) in (b), 
and DM (YMW16) in (c).
Both measurements with real data (red) and Monte Carlo realizations (black) are presented.} 
\label{fig: sf}
\end{figure*}

Fig. \ref{fig: sf} presents the $D(\theta)$ of the subset of FRBs, with the total DM used in Fig. \ref{fig: sftot}, 
and the DM excess between DM determined by {\it fitburst}
\citep{chim21}
and Milky Way DM modeled by 
NE2001 
\citep{Cor02}, i.e., DM (NE2001),
in Fig. \ref{fig: sfne}
and 
YMW16
\citep{Yao17}, i.e., DM (YMW16),
in Fig. \ref{fig: sfymw}.
We see that $D(\theta)$ is not sensitive to the modeled Milky Way DMs. 
In all cases, it shows 
{\xu non-flat $D(\theta)$ at small $\theta$.}
Non-flat SF indicates possible correlation of electron density fluctuations within scales 
$\approx 2000~\text{Mpc}\times 10^\circ  \approx
350$ Mpc.
To evaluate the statistical uncertainty, we perform Monte Carlo simulations by 
randomizing DMs while holding the locations of FRBs fixed. 
As shown by $D(\theta)$ measured 
from $40$ Monte Carlo realizations,
the uncertainty becomes larger 
toward smaller $\theta$ with 
fewer pairs of FRBs available.

\begin{figure*}[htbp]
\centering   
\subfigure[Total detection DM]{
   \includegraphics[width=8.5cm]{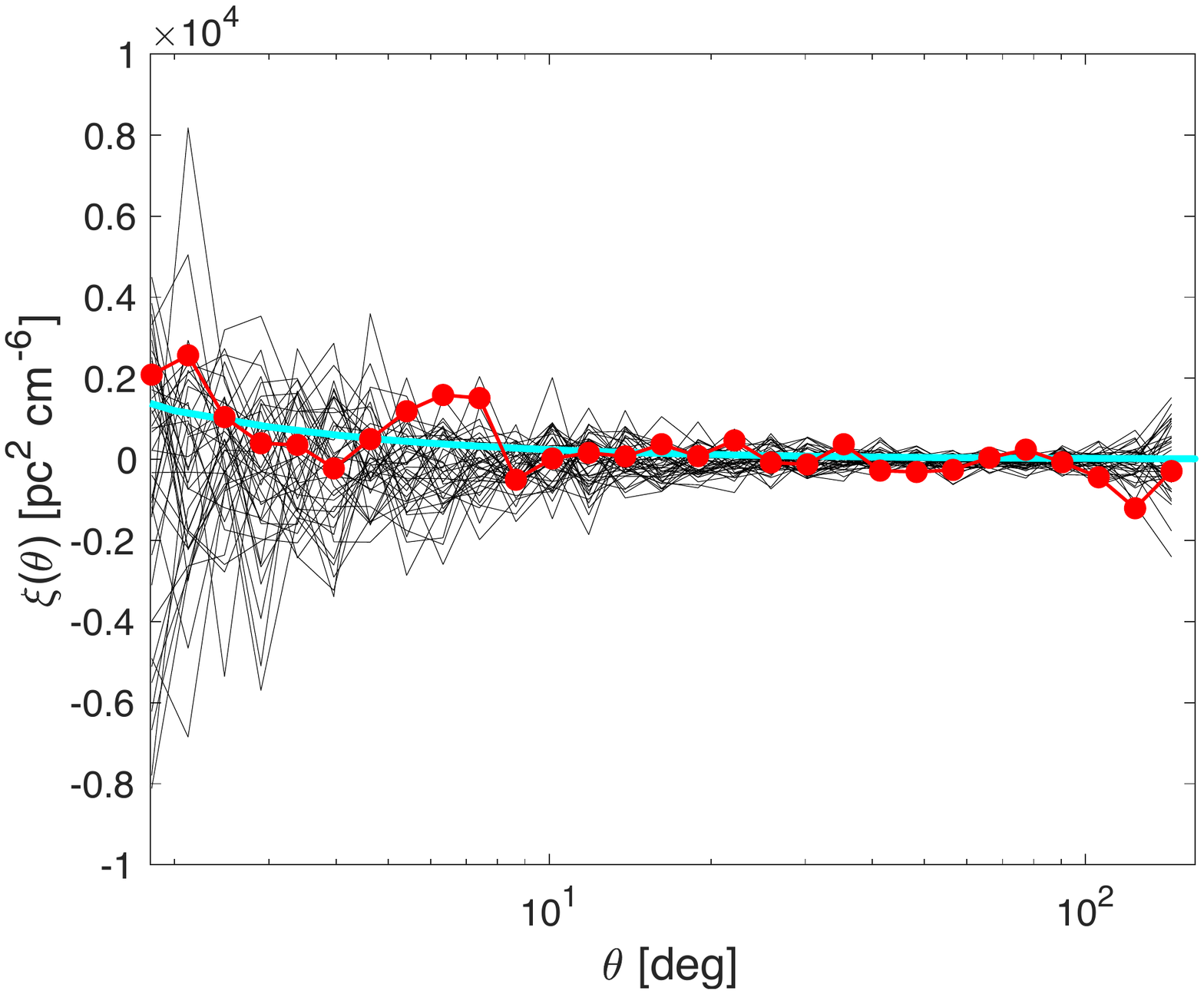}\label{fig: cftot}}
\subfigure[DM (NE2001)]{
   \includegraphics[width=8.5cm]{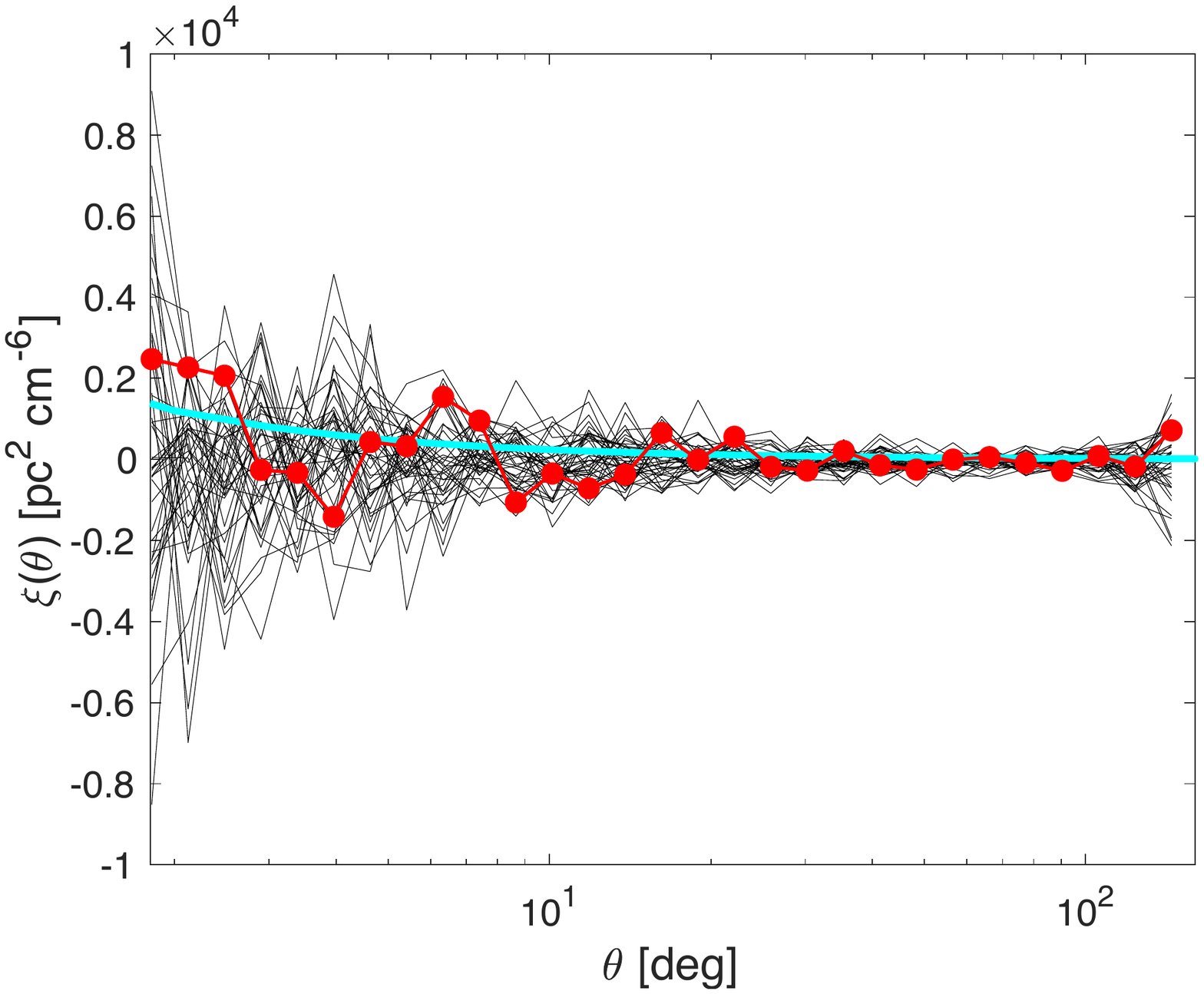}\label{fig: cfne}}
\subfigure[DM (YMW16)]{
   \includegraphics[width=8.5cm]{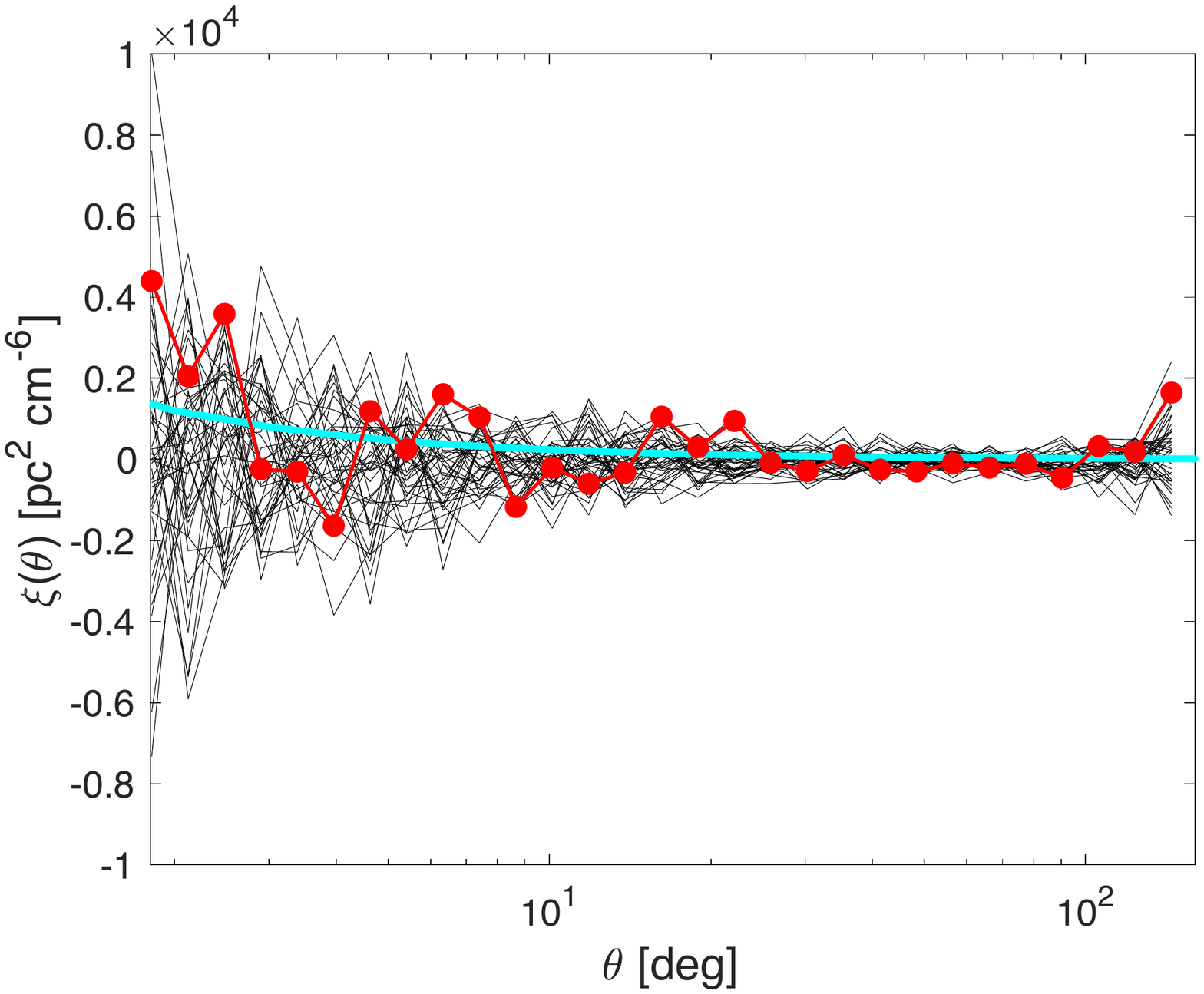}\label{fig: cfymw}}
\caption{Same as Fig. \ref{fig: sf} but for $\xi(\theta)$. 
The cyan line shows the analytical approximation based on cosmological simulations
given by 
\citet{Takaha21}.
}
\label{fig: cf}
\end{figure*}

The measured $\xi(\theta)$ is shown in Fig. \ref{fig: cf}.
It is more sensitive to the modeled Milky Way DMs than $D(\theta)$. 
For comparison, we also present 
the analytically modeled $\xi(\theta;z_s)$ by 
\citet{Takaha21}.
In their model
the free-electron abundance and the power spectrum of its spatial fluctuations are measured from hydrodynamic cosmological simulations IllustrisTNG300
\citep{Nelson18}.
The cyan line shows its 
analytical approximation at $\theta \gtrsim 1^\circ$, 
\begin{equation}
    \xi(\theta;z_s) \approx 2400 \Big(\frac{\theta}{\text{deg}}\Big)^{-1} \text{pc}^2\text{cm}^{-6},
\end{equation}
where $z_s$ is the source redshift. 
\citet{Takaha21} 
found that the above expression is insensitive to $z_s$ at $z_s \gtrsim 0.3$ as a large-scale signal is dominated by nearby structures. 
This further justifies our use of a subset of FRBs with relatively small DMs.
Our measured $\xi(\theta)$, with a large uncertainty, shows some trend of increasing $\xi(\theta)$ toward smaller $\theta$ at $\theta<10^\circ$. 
Toward larger $\theta$, 
$\xi(\theta)$ gradually approaches zero as theoretically expected. 
We see that as the theoretically modeled $\xi(\theta;z_s)$ is small, 
a more accurate comparison between the model and observational measurements  
requires a larger sample size and a higher angular resolution.

\begin{figure*}[htbp]
\centering   
\subfigure[Total detection DM]{
   \includegraphics[width=8.5cm]{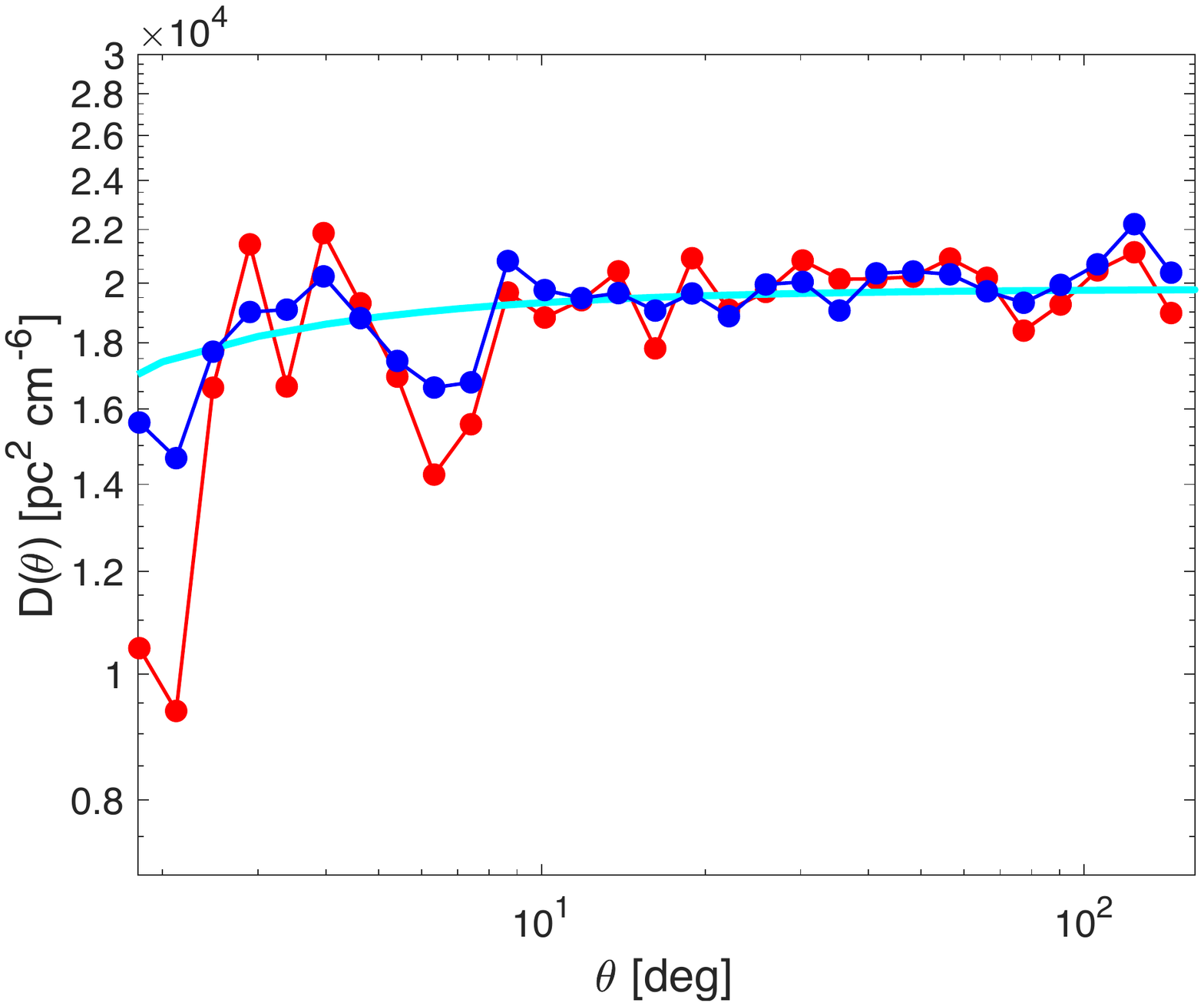}\label{fig: cfsftot}}
\subfigure[DM (NE2001)]{
   \includegraphics[width=8.5cm]{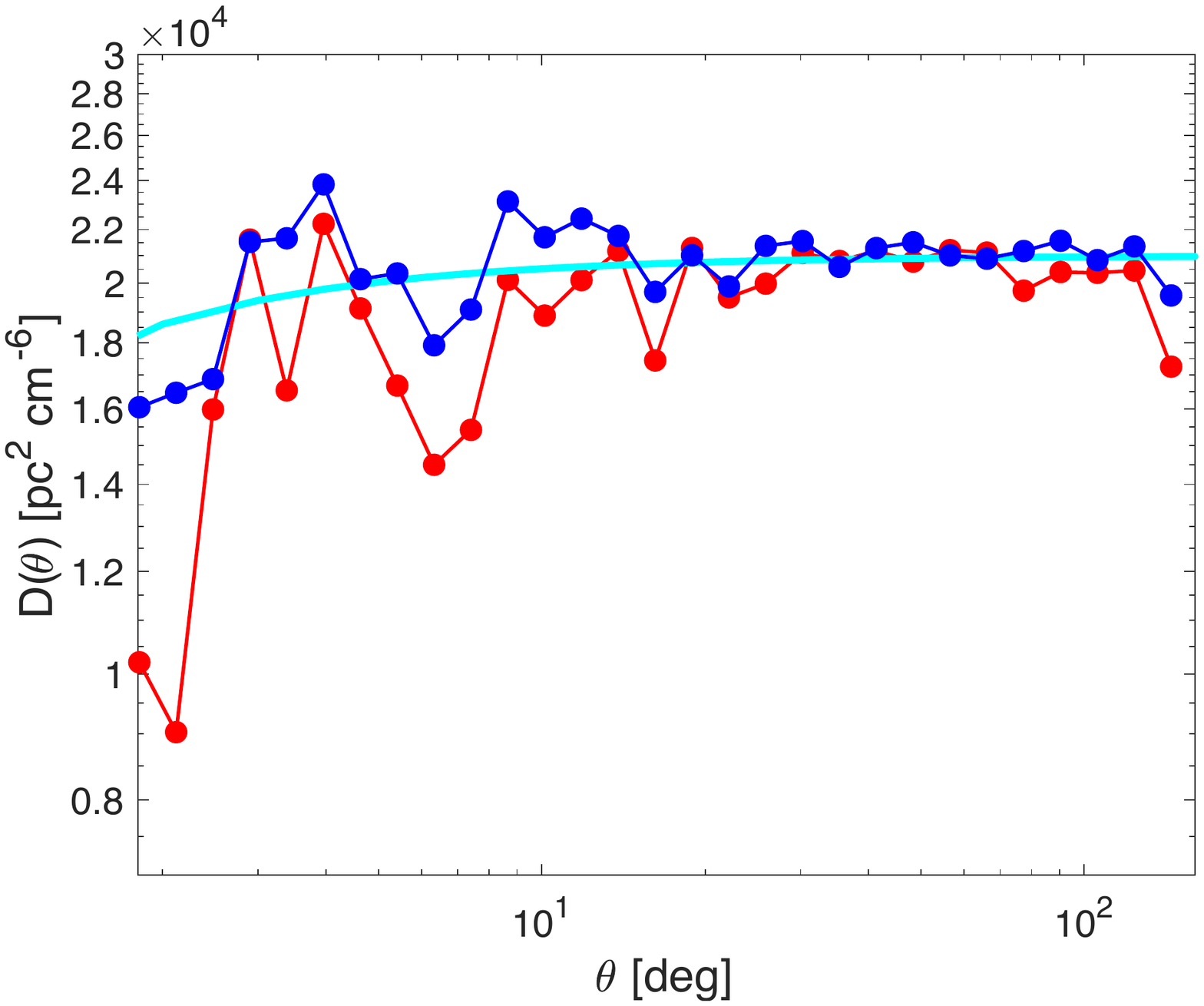}\label{fig: cfsfne}}
\subfigure[DM (YMW16)]{
   \includegraphics[width=8.5cm]{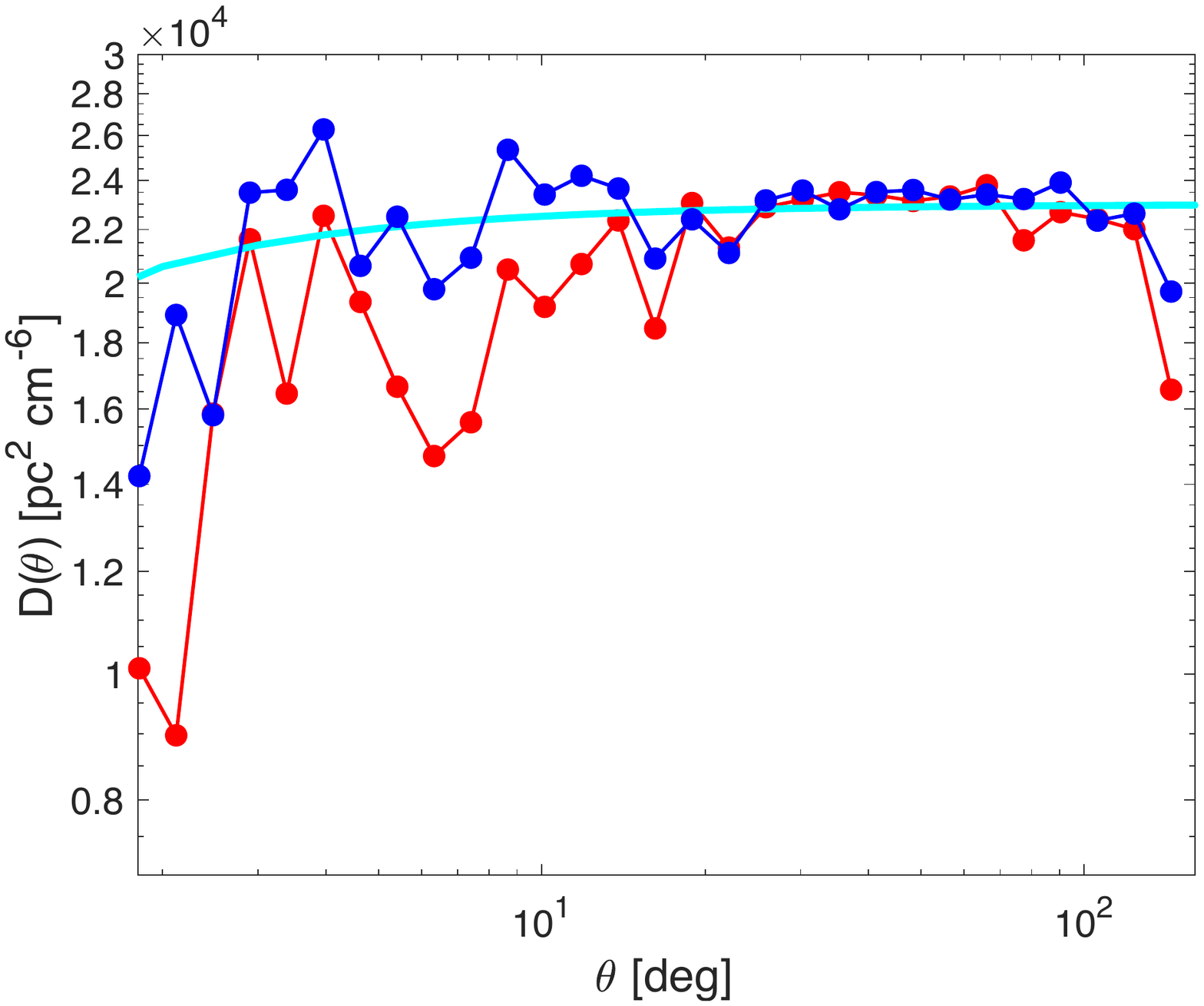}\label{fig: cfsfymw}}
\caption{Comparison between directly measured $D(\theta)$ (red) and $D(\theta)$ derived from measured $\xi(\theta)$ (dark blue).
The cyan line corresponds to 
$D(\theta)$ derived from the analytical approximation of $\xi(\theta;z_s)$ given by 
\citet{Takaha21}.}
\label{fig: cfsf}
\end{figure*}

To examine whether $D(\theta)$ and $\xi(\theta)$ of FRBs provide similar information on the correlation of electron density fluctuations, 
in Fig. \ref{fig: cfsf},
we compare the measured $D(\theta)$ with that derived from $\xi(\theta)$ by using the relation in Eq. \eqref{eq:sfcf}.
The constant value in 
Eq. \eqref{eq:sfcf} is determined by visually matching the two at large $\theta$, 
which is 
$2.0\times10^4$ 
pc$^2$ cm$^{-6}$ for the case with total DMs, 
and slightly larger for modeled extragalactic DMs, i.e., 
$2.1\times10^4$ pc$^2$ cm$^{-6}$ for DM (NE2001) and 
$2.3\times10^4$ pc$^2$ cm$^{-6}$
for DM (YMW16).
We see that the
measured $D(\theta)$ and that derived from $\xi(\theta)$ have a better agreement at $\theta \gtrsim 10^\circ$, 
indicative of the homogeneity of the large-scale density distribution.
The discrepancy seen at smaller $\theta$ 
can be caused by the limited sample size, as smaller-$\theta$ measurements suffer larger statistical uncertainties. 
It can also be caused by intermediate-scale
(on the order of $100$ Mpc) inhomogeneities of intergalactic density distribution.
Possible inhomogeneities in the universe on such length scales were suggested in e.g., \citet{Kop88,Sylo09,Sylos11,Perivo14},
based on the
observed galaxy distribution and other effects,
though these claims are controversial
(e.g., \citealt{Hogg2005,Ntelis2017}).
In the latter situation, 
we do not expect that 
$D(\theta)$ and $\xi(\theta)$ 
can be determined from each other at small $\theta$
even with a larger sample of FRBs. 
As the SF is less sensitive to large-scale fluctuations and has a higher accuracy using less data compared with the CF 
\citep{Schul81},
the statistical measurements of DM fluctuations at small $\theta$
with SF can be more reliable and informative.

\begin{figure*}[htbp]
\centering   
\subfigure[]{
   \includegraphics[width=8.5cm]{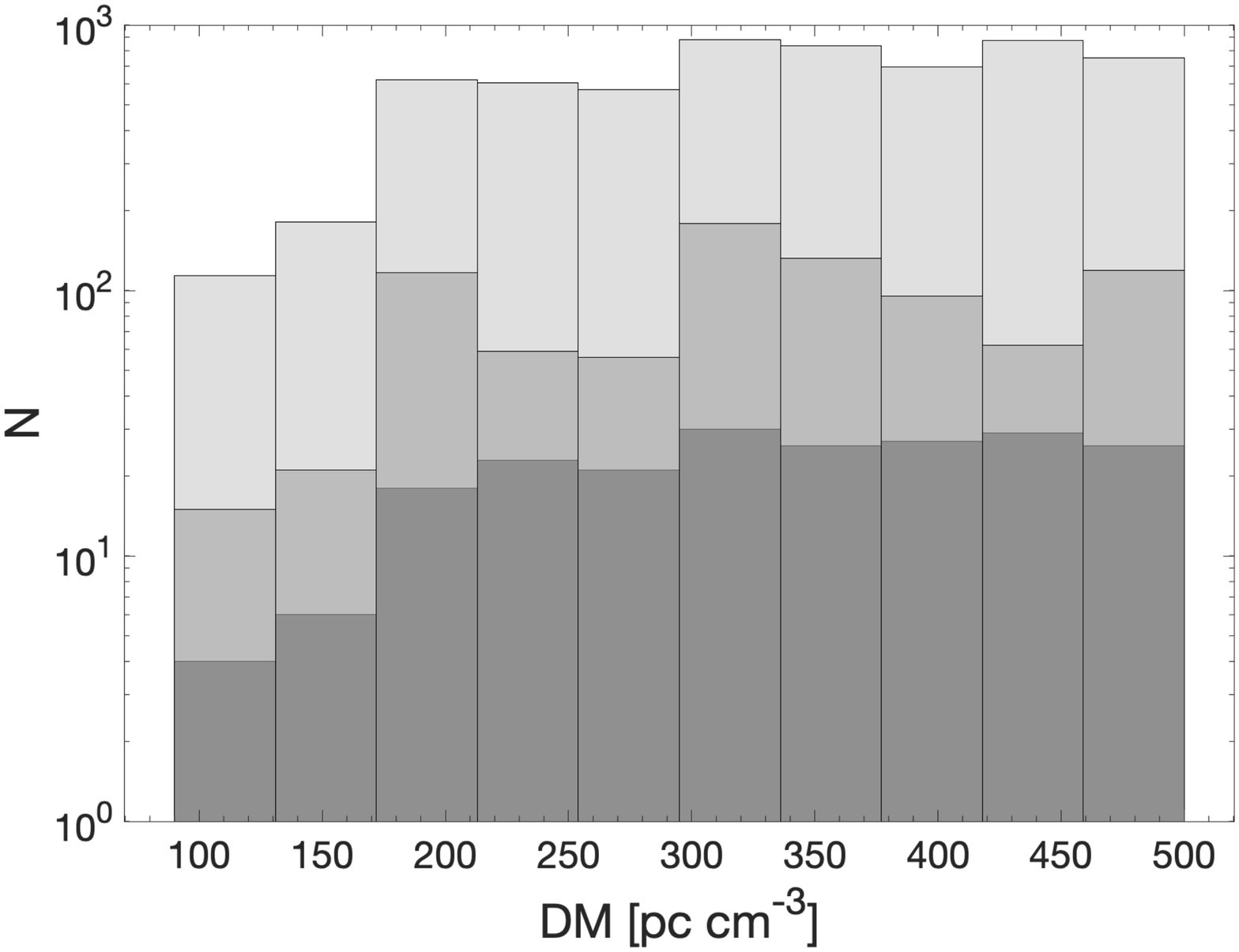}\label{fig: rejpdf}}
   
\subfigure[$855$ mock FRBs]{
   \includegraphics[width=8.5cm]{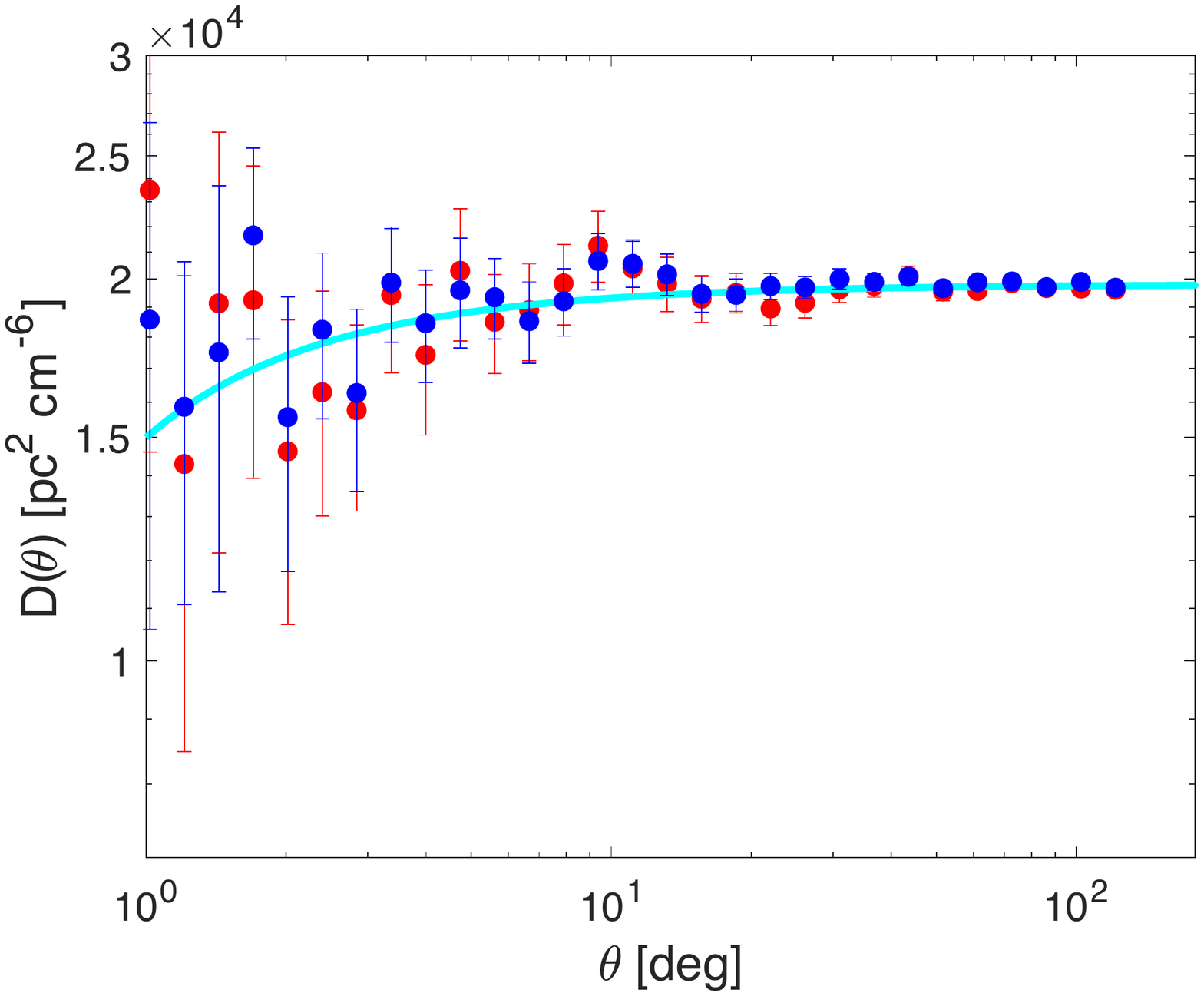}\label{fig: 855sfcf}}
\subfigure[$6136$ mock FRBs]{
   \includegraphics[width=8.5cm]{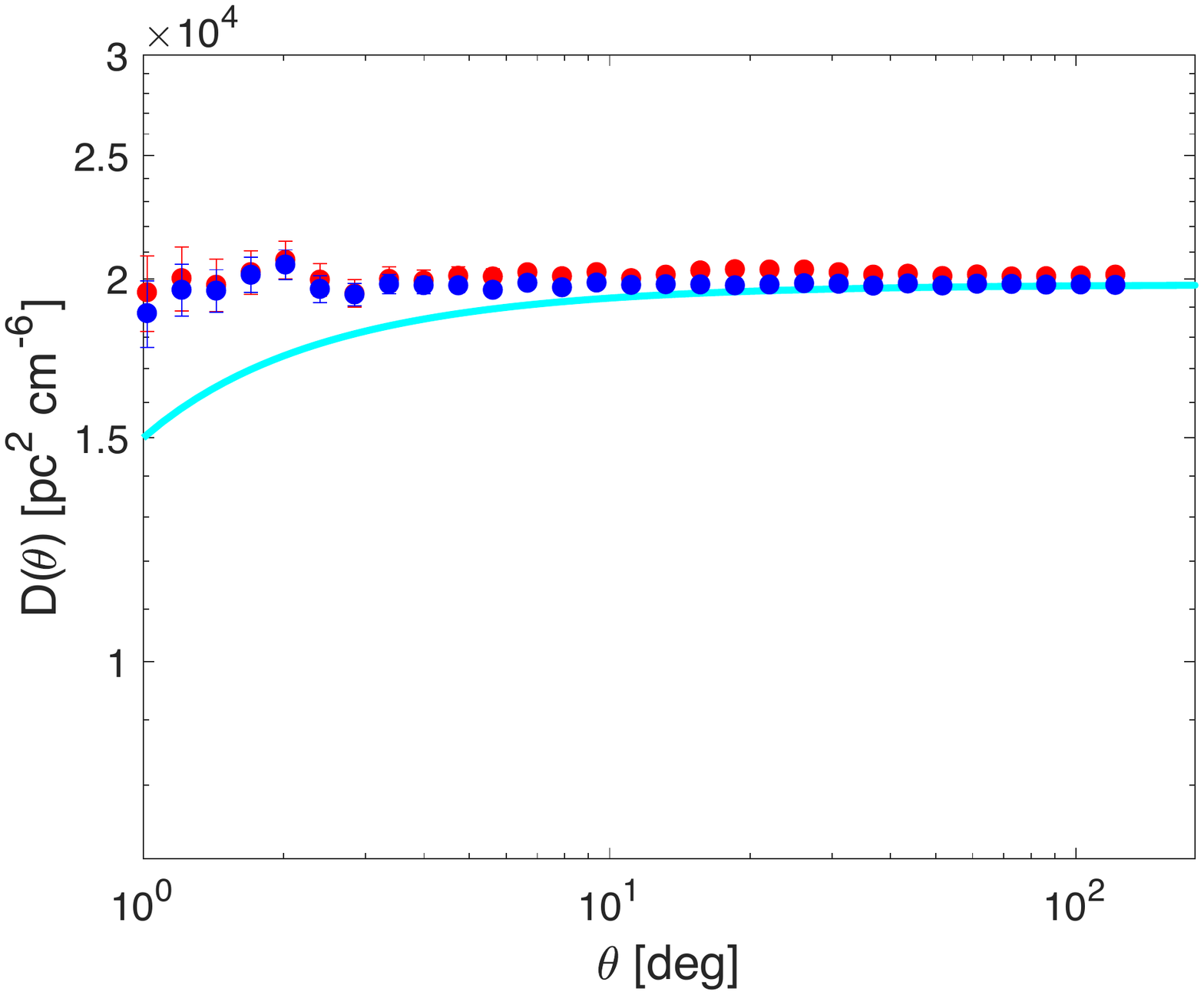}\label{fig: 6136sfcf}}
\caption{\xu (a) DM distributions of 
CHIME FRBs with DM$<500$ pc cm$^{-3}$ (dark gray) and
mock samples of $855$ (gray)
and $6136$ (light gray) FRBs.
(b) and (c) Same as Fig. \ref{fig: cfsf} 
but for DMs of mock FRBs. 
Error bars indicate $95\%$ confidence intervals.
}
\label{fig: recsam}
\end{figure*}

{\xu Obviously, 
the quantitative comparison between the 
simulated and observationally measured statistics of DM fluctuations requires a sufficiently large sample of FRBs. 
By using the rejection sampling method 
\citep{Mack03},
we generate a large sample of mock FRBs from the target DM distribution of the CHIME FRBs with DM$<500$ pc cm$^{-3}$ (see Fig. \ref{fig: rejpdf}). 
Under the consideration that future radio telescopes, such as the Square Kilometer Array, 
will enable full sky coverage,  
we have the mock FRBs randomly distributed across the entire sky. 
The measured $D(\theta)$ and that derived from $\xi(\theta)$ of the mock samples are presented in Figs. \ref{fig: 855sfcf} and \ref{fig: 6136sfcf}.
It shows that 
at least several thousands of FRBs are needed to have a clean comparison
with the expectation from cosmological simulations. 
A larger angular resolution is achieved with the increase of sample size. 
We note that 
as the DM fluctuations of mock FRBs 
are uncorrelated and homogeneous, 
the flatness of $D(\theta)$ and the good match between $D(\theta)$ and that converted from $\xi(\theta)$ are expected for a sufficiently large sample 
(see Fig. \ref{fig: 6136sfcf}).
}

\section{Summary}

We compared the SF of DM fluctuations measured in XZ20 with FRBCAT
and that measured with a larger sample of FRBs from the First CHIME/FRB Catalog. 
We found consistent DM fluctuation level at $\theta\gtrsim 10^\circ$ in both cases, 
but no clear {\xu power-law trend} at smaller $\theta$ hinting towards large scale turbulence can be recovered with the larger sample.
The apparent signal in the earlier FRBCAT analysis is likely to
be a statistical fluctuation caused by the scarcity
of close FRB pairs in the
FRBCAT sample, 
but the effect of different sky coverage of the two catalogs cannot be completely excluded.

To suppress the
distortions by distances and host DMs on the statistical properties of 
intergalactic electron density fluctuations, 
we introduced a tentative DM cut
and focused on a subset of CHIME FRBs with DM$<500$ pc cm$^{-3}$.
{\xu Non-flat} $D(\theta)$ is observed at $\theta<10^\circ$ for the subsample, 
but the statistical uncertainty is large due to the DM cut. 

We also measured the CF, i.e.,  $\xi(\theta)$,
of the subsample. 
Compared with $D(\theta)$, 
it is more sensitive to the Milky Way DM models, 
showing more obvious 
differences between the measurements with 
total DMs and modeled extragalactic DMs.
A trend of increasing $\xi(\theta)$ with decreasing $\theta$ is seen, with a large statistical uncertainty. 
As the theoretically expected correlation signal is weak
\citep{Takaha21},
a larger sample size with a higher
angular resolution is necessary for a more accurate comparison 
between the modeled and observationally measured $\xi(\theta)$.

The 
discrepancy between $D(\theta)$ and
{\xu that converted from $\xi(\theta)$} at $\theta \lesssim 10^\circ$ 
can arise from large statistical uncertainties or density inhomogeneities on scales on the order of $100$ Mpc. 
In the latter situation, the discrepancy would persist when more FRBs are available. 
Its potentially important cosmological implications on, e.g., determination of the Hubble constant
\citep{Fani21},
deserve further study. 
{\xu The inhomogeneous distribution of matter on scales $\lesssim 100$ Mpc
can give rise to deviations from isotropic and homogeneous expansion due to gravitational induced peculiar velocities, and thus the value of the Hubble constant measured at distances $\lesssim 100$ Mpc has a large scatter
(e.g., \citealt{Shi98,Free03}).}
{\xu As demonstrated by our analysis with mock FRBs,  
a large sample of several thousands of FRBs is needed for 
a clean comparison with the expectation from cosmological simulations 
and 
determining the effect of statistical uncertainties on the apparent disagreement between $D(\theta)$ and that converted from $\xi(\theta)$ of the current FRB sample.}

\acknowledgments
S.X. acknowledges the support for 
this work provided by NASA through the NASA Hubble Fellowship grant \# HST-HF2-51473.001-A awarded by the Space Telescope Science Institute, which is operated by the Association of Universities for Research in Astronomy, Incorporated, under NASA contract NAS5-26555. 
\software{MATLAB \citep{MATLAB:2018}}

\bibliographystyle{aasjournal}
\bibliography{xu}

\begin{thebibliography}{}
\expandafter\ifx\csname natexlab\endcsname\relax\def\natexlab#1{#1}\fi
\providecommand{\url}[1]{\href{#1}{#1}}
\providecommand{\dodoi}[1]{doi:~\href{http://doi.org/#1}{\nolinkurl{#1}}}
\providecommand{\doeprint}[1]{\href{http://ascl.net/#1}{\nolinkurl{http://ascl.net/#1}}}
\providecommand{\doarXiv}[1]{\href{https://arxiv.org/abs/#1}{\nolinkurl{https://arxiv.org/abs/#1}}}

\bibitem[{{Bahcall} \& {Burgett}(1986)}]{Bah86}
{Bahcall}, N.~A., \& {Burgett}, W.~S. 1986, \apjl, 300, L35,
  \dodoi{10.1086/184598}

\bibitem[{{Beniamini} {et~al.}(2021){Beniamini}, {Kumar}, {Ma}, \&
  {Quataert}}]{Beniamini21}
{Beniamini}, P., {Kumar}, P., {Ma}, X., \& {Quataert}, E. 2021, \mnras, 502,
  5134, \dodoi{10.1093/mnras/stab309}

\bibitem[{{Caleb} {et~al.}(2019){Caleb}, {Flynn}, \& {Stappers}}]{Caleb19}
{Caleb}, M., {Flynn}, C., \& {Stappers}, B.~W. 2019, \mnras, 485, 2281,
  \dodoi{10.1093/mnras/stz571}

\bibitem[{{Cordes} \& {Chatterjee}(2019)}]{Cord19}
{Cordes}, J.~M., \& {Chatterjee}, S. 2019, \araa, 57, 417,
  \dodoi{10.1146/annurev-astro-091918-104501}

\bibitem[{{Cordes} \& {Lazio}(2002)}]{Cor02}
{Cordes}, J.~M., \& {Lazio}, T.~J.~W. 2002, ArXiv Astrophysics e-print:
  astro-ph/0207156

\bibitem[{{Deng} \& {Zhang}(2014)}]{Deng14}
{Deng}, W., \& {Zhang}, B. 2014, \apjl, 783, L35,
  \dodoi{10.1088/2041-8205/783/2/L35}

\bibitem[{{Einasto} {et~al.}(2020){Einasto}, {H{\"u}tsi}, \&
  {Einasto}}]{Eina20}
{Einasto}, J., {H{\"u}tsi}, G., \& {Einasto}, M. 2020, arXiv e-prints,
  arXiv:2004.03232.
\newblock \doarXiv{2004.03232}

\bibitem[{{Fanizza} {et~al.}(2021){Fanizza}, {Fiorini}, \& {Marozzi}}]{Fani21}
{Fanizza}, G., {Fiorini}, B., \& {Marozzi}, G. 2021, arXiv e-prints,
  arXiv:2102.12419.
\newblock \doarXiv{2102.12419}

\bibitem[{{Freedman} \& {Turner}(2003)}]{Free03}
{Freedman}, W.~L., \& {Turner}, M.~S. 2003, Reviews of Modern Physics, 75,
  1433, \dodoi{10.1103/RevModPhys.75.1433}

\bibitem[{{Gao} {et~al.}(2014){Gao}, {Li}, \& {Zhang}}]{Gao14}
{Gao}, H., {Li}, Z., \& {Zhang}, B. 2014, \apj, 788, 189,
  \dodoi{10.1088/0004-637X/788/2/189}

\bibitem[{{Ha} {et~al.}(2021){Ha}, {Li}, {Xu}, {Kounkel}, \& {Li}}]{Ha21}
{Ha}, T., {Li}, Y., {Xu}, S., {Kounkel}, M., \& {Li}, H. 2021, \apjl, 907, L40,
  \dodoi{10.3847/2041-8213/abd8c9}

\bibitem[{{Hawkins} {et~al.}(2003){Hawkins}, {Maddox}, {Cole}, {Lahav},
  {Madgwick}, {Norberg}, {Peacock}, {Baldry}, {Baugh}, {Bland-Hawthorn},
  {Bridges}, {Cannon}, {Colless}, {Collins}, {Couch}, {Dalton}, {De Propris},
  {Driver}, {Efstathiou}, {Ellis}, {Frenk}, {Glazebrook}, {Jackson}, {Jones},
  {Lewis}, {Lumsden}, {Percival}, {Peterson}, {Sutherland}, \&
  {Taylor}}]{Hawk03}
{Hawkins}, E., {Maddox}, S., {Cole}, S., {et~al.} 2003, \mnras, 346, 78,
  \dodoi{10.1046/j.1365-2966.2003.07063.x}

\bibitem[{{Hogg} {et~al.}(2005){Hogg}, {Eisenstein}, {Blanton}, {Bahcall},
  {Brinkmann}, {Gunn}, \& {Schneider}}]{Hogg2005}
{Hogg}, D.~W., {Eisenstein}, D.~J., {Blanton}, M.~R., {et~al.} 2005, \apj, 624,
  54, \dodoi{10.1086/429084}

\bibitem[{{Keane} {et~al.}(2016){Keane}, {Johnston}, {Bhandari}, {Barr},
  {Bhat}, {Burgay}, {Caleb}, {Flynn}, {Jameson}, {Kramer}, {Petroff},
  {Possenti}, {van Straten}, {Bailes}, {Burke-Spolaor}, {Eatough}, {Stappers},
  {Totani}, {Honma}, {Furusawa}, {Hattori}, {Morokuma}, {Niino}, {Sugai},
  {Terai}, {Tominaga}, {Yamasaki}, {Yasuda}, {Allen}, {Cooke}, {Jencson},
  {Kasliwal}, {Kaplan}, {Tingay}, {Williams}, {Wayth}, {Chandra}, {Perrodin},
  {Berezina}, {Mickaliger}, \& {Bassa}}]{Keane16}
{Keane}, E.~F., {Johnston}, S., {Bhandari}, S., {et~al.} 2016, \nat, 530, 453,
  \dodoi{10.1038/nature17140}

\bibitem[{{Klypin} \& {Kopylov}(1983)}]{Klyp83}
{Klypin}, A.~A., \& {Kopylov}, A.~I. 1983, Pisma v Astronomicheskii Zhurnal, 9,
  75

\bibitem[{{Kopylov} {et~al.}(1988){Kopylov}, {Kuznetsov}, {Fetisova}, \&
  {Shvartsman}}]{Kop88}
{Kopylov}, A.~I., {Kuznetsov}, D.~Y., {Fetisova}, T.~S., \& {Shvartsman}, V.~F.
  1988, in Large Scale Structures of the Universe, ed. J.~{Audouze}, M.~C.
  {Pelletan}, A.~{Szalay}, Y.~B. {Zel'dovich}, \& P.~J.~E. {Peebles}, Vol. 130,
  129

\bibitem[{{Kumar} \& {Linder}(2019)}]{Kumar19}
{Kumar}, P., \& {Linder}, E.~V. 2019, \prd, 100, 083533,
  \dodoi{10.1103/PhysRevD.100.083533}

\bibitem[{{Landy} \& {Szalay}(1993)}]{LanSza93}
{Landy}, S.~D., \& {Szalay}, A.~S. 1993, \apj, 412, 64, \dodoi{10.1086/172900}

\bibitem[{{Lazarian} \& {Pogosyan}(2016)}]{LP16}
{Lazarian}, A., \& {Pogosyan}, D. 2016, \apj, 818, 178,
  \dodoi{10.3847/0004-637X/818/2/178}

\bibitem[{{Li} {et~al.}(2020){Li}, {Gendron-Marsolais}, {Zhuravleva}, {Xu},
  {Simionescu}, {Tremblay}, {Lochhaas}, {Bryan}, {Quataert}, {Murray},
  {Boselli}, {Hlavacek-Larrondo}, {Zheng}, {Fossati}, {Li}, {Emsellem},
  {Sarzi}, {Arzamasskiy}, \& {Vishniac}}]{Li20}
{Li}, Y., {Gendron-Marsolais}, M.-L., {Zhuravleva}, I., {et~al.} 2020, \apjl,
  889, L1, \dodoi{10.3847/2041-8213/ab65c7}

\bibitem[{{Limber}(1953)}]{Limber53}
{Limber}, D.~N. 1953, \apj, 117, 134, \dodoi{10.1086/145672}

\bibitem[{{Lorimer} {et~al.}(2007){Lorimer}, {Bailes}, {McLaughlin},
  {Narkevic}, \& {Crawford}}]{Lor07}
{Lorimer}, D.~R., {Bailes}, M., {McLaughlin}, M.~A., {Narkevic}, D.~J., \&
  {Crawford}, F. 2007, Science, 318, 777, \dodoi{10.1126/science.1147532}

\bibitem[{{Mackay}(2003)}]{Mack03}
{Mackay}, D. J.~C. 2003, {Information Theory, Inference and Learning
  Algorithms}

\bibitem[{{Macquart} \& {Koay}(2013)}]{Macq13}
{Macquart}, J.-P., \& {Koay}, J.~Y. 2013, \apj, 776, 125,
  \dodoi{10.1088/0004-637X/776/2/125}

\bibitem[{{Macquart} {et~al.}(2020){Macquart}, {Prochaska}, {McQuinn},
  {Bannister}, {Bhandari}, {Day}, {Deller}, {Ekers}, {James}, {Marnoch},
  {Os{\l}owski}, {Phillips}, {Ryder}, {Scott}, {Shannon}, \&
  {Tejos}}]{Macqnat20}
{Macquart}, J.~P., {Prochaska}, J.~X., {McQuinn}, M., {et~al.} 2020, \nat, 581,
  391, \dodoi{10.1038/s41586-020-2300-2}

\bibitem[{{Masui} \& {Sigurdson}(2015)}]{Masui15}
{Masui}, K.~W., \& {Sigurdson}, K. 2015, \prl, 115, 121301,
  \dodoi{10.1103/PhysRevLett.115.121301}

\bibitem[{MATLAB(2018)}]{MATLAB:2018}
MATLAB. 2018, 9.7.0.1190202 (R2019b) (Natick, Massachusetts: The MathWorks
  Inc.)

\bibitem[{Monin \& Yaglom(1965)}]{monin1965}
Monin, A., \& Yaglom, A. 1965, Statistical Fluid Mechanics, Volume II:
  Mechanics of Turbulence (Nauka Press, Moscow)

\bibitem[{{Nelson} {et~al.}(2018){Nelson}, {Pillepich}, {Springel},
  {Weinberger}, {Hernquist}, {Pakmor}, {Genel}, {Torrey}, {Vogelsberger},
  {Kauffmann}, {Marinacci}, \& {Naiman}}]{Nelson18}
{Nelson}, D., {Pillepich}, A., {Springel}, V., {et~al.} 2018, \mnras, 475, 624,
  \dodoi{10.1093/mnras/stx3040}

\bibitem[{{Niu} {et~al.}(2021){Niu}, {Aggarwal}, {Li}, {Zhang}, {Chatterjee},
  {Tsai}, {Yu}, {Law}, {Burke-Spolaor}, {Cordes}, {Zhang}, {Ocker}, {Yao},
  {Wang}, {Feng}, {Niino}, {Bochenek}, {Cruces}, {Connor}, {Jiang}, {Dai},
  {Luo}, {Li}, {Miao}, {Niu}, {Anna-Thomas}, {Sydnor}, {Stern}, {Wang}, {Yuan},
  {Yue}, {Zhou}, {Yan}, {Zhu}, \& {Zhang}}]{Niu21}
{Niu}, C.~H., {Aggarwal}, K., {Li}, D., {et~al.} 2021, arXiv e-prints,
  arXiv:2110.07418.
\newblock \doarXiv{2110.07418}

\bibitem[{{Ntelis} {et~al.}(2017){Ntelis}, {Hamilton}, {Le Goff}, {Burtin},
  {Laurent}, {Rich}, {Guillermo Busca}, {Tinker}, {Aubourg}, {du Mas des
  Bourboux}, {Bautista}, {Palanque Delabrouille}, {Delubac}, {Eftekharzadeh},
  {Hogg}, {Myers}, {Vargas-Maga{\~n}a}, {P{\^a}ris}, {Petitjean}, {Rossi},
  {Schneider}, {Tojeiro}, \& {Yeche}}]{Ntelis2017}
{Ntelis}, P., {Hamilton}, J.-C., {Le Goff}, J.-M., {et~al.} 2017, \jcap, 2017,
  019, \dodoi{10.1088/1475-7516/2017/06/019}

\bibitem[{{Peebles}(1980)}]{Peeb80}
{Peebles}, P.~J.~E. 1980, {The large-scale structure of the universe}

\bibitem[{{Perivolaropoulos}(2014)}]{Perivo14}
{Perivolaropoulos}, L. 2014, Galaxies, 2, 22, \dodoi{10.3390/galaxies2010022}

\bibitem[{{Petroff} {et~al.}(2016){Petroff}, {Barr}, {Jameson}, {Keane},
  {Bailes}, {Kramer}, {Morello}, {Tabbara}, \& {van Straten}}]{Pat16}
{Petroff}, E., {Barr}, E.~D., {Jameson}, A., {et~al.} 2016, PASA, 33, e045,
  \dodoi{10.1017/pasa.2016.35}

\bibitem[{{Planck Collaboration} {et~al.}(2016){Planck Collaboration}, {Ade},
  {Aghanim}, {Arnaud}, {Ashdown}, {Aumont}, {Baccigalupi}, {Banday},
  {Barreiro}, {Bartlett}, {Bartolo}, {Battaner}, {Battye}, {Benabed},
  {Beno{\^\i}t}, {Benoit-L{\'e}vy}, {Bernard}, {Bersanelli}, {Bielewicz},
  {Bock}, {Bonaldi}, {Bonavera}, {Bond}, {Borrill}, {Bouchet}, {Boulanger},
  {Bucher}, {Burigana}, {Butler}, {Calabrese}, {Cardoso}, {Catalano},
  {Challinor}, {Chamballu}, {Chary}, {Chiang}, {Chluba}, {Christensen},
  {Church}, {Clements}, {Colombi}, {Colombo}, {Combet}, {Coulais}, {Crill},
  {Curto}, {Cuttaia}, {Danese}, {Davies}, {Davis}, {de Bernardis}, {de Rosa},
  {de Zotti}, {Delabrouille}, {D{\'e}sert}, {Di Valentino}, {Dickinson},
  {Diego}, {Dolag}, {Dole}, {Donzelli}, {Dor{\'e}}, {Douspis}, {Ducout},
  {Dunkley}, {Dupac}, {Efstathiou}, {Elsner}, {En{\ss}lin}, {Eriksen},
  {Farhang}, {Fergusson}, {Finelli}, {Forni}, {Frailis}, {Fraisse},
  {Franceschi}, {Frejsel}, {Galeotta}, {Galli}, {Ganga}, {Gauthier}, {Gerbino},
  {Ghosh}, {Giard}, {Giraud-H{\'e}raud}, {Giusarma}, {Gjerl{\o}w},
  {Gonz{\'a}lez-Nuevo}, {G{\'o}rski}, {Gratton}, {Gregorio}, {Gruppuso},
  {Gudmundsson}, {Hamann}, {Hansen}, {Hanson}, {Harrison}, {Helou},
  {Henrot-Versill{\'e}}, {Hern{\'a}ndez-Monteagudo}, {Herranz}, {Hildebrand t},
  {Hivon}, {Hobson}, {Holmes}, {Hornstrup}, {Hovest}, {Huang}, {Huffenberger},
  {Hurier}, {Jaffe}, {Jaffe}, {Jones}, {Juvela}, {Keih{\"a}nen}, {Keskitalo},
  {Kisner}, {Kneissl}, {Knoche}, {Knox}, {Kunz}, {Kurki-Suonio}, {Lagache},
  {L{\"a}hteenm{\"a}ki}, {Lamarre}, {Lasenby}, {Lattanzi}, {Lawrence}, {Leahy},
  {Leonardi}, {Lesgourgues}, {Levrier}, {Lewis}, {Liguori}, {Lilje},
  {Linden-V{\o}rnle}, {L{\'o}pez-Caniego}, {Lubin}, {Mac{\'\i}as-P{\'e}rez},
  {Maggio}, {Maino}, {Mandolesi}, {Mangilli}, {Marchini}, {Maris}, {Martin},
  {Martinelli}, {Mart{\'\i}nez-Gonz{\'a}lez}, {Masi}, {Matarrese}, {McGehee},
  {Meinhold}, {Melchiorri}, {Melin}, {Mendes}, {Mennella}, {Migliaccio},
  {Millea}, {Mitra}, {Miville-Desch{\^e}nes}, {Moneti}, {Montier}, {Morgante},
  {Mortlock}, {Moss}, {Munshi}, {Murphy}, {Naselsky}, {Nati}, {Natoli},
  {Netterfield}, {N{\o}rgaard-Nielsen}, {Noviello}, {Novikov}, {Novikov},
  {Oxborrow}, {Paci}, {Pagano}, {Pajot}, {Paladini}, {Paoletti}, {Partridge},
  {Pasian}, {Patanchon}, {Pearson}, {Perdereau}, {Perotto}, {Perrotta},
  {Pettorino}, {Piacentini}, {Piat}, {Pierpaoli}, {Pietrobon}, {Plaszczynski},
  {Pointecouteau}, {Polenta}, {Popa}, {Pratt}, {Pr{\'e}zeau}, {Prunet},
  {Puget}, {Rachen}, {Reach}, {Rebolo}, {Reinecke}, {Remazeilles}, {Renault},
  {Renzi}, {Ristorcelli}, {Rocha}, {Rosset}, {Rossetti}, {Roudier},
  {Rouill{\'e} d'Orfeuil}, {Rowan-Robinson}, {Rubi{\~n}o-Mart{\'\i}n},
  {Rusholme}, {Said}, {Salvatelli}, {Salvati}, {Sandri}, {Santos},
  {Savelainen}, {Savini}, {Scott}, {Seiffert}, {Serra}, {Shellard}, {Spencer},
  {Spinelli}, {Stolyarov}, {Stompor}, {Sudiwala}, {Sunyaev}, {Sutton},
  {Suur-Uski}, {Sygnet}, {Tauber}, {Terenzi}, {Toffolatti}, {Tomasi},
  {Tristram}, {Trombetti}, {Tucci}, {Tuovinen}, {T{\"u}rler}, {Umana},
  {Valenziano}, {Valiviita}, {Van Tent}, {Vielva}, {Villa}, {Wade}, {Wandelt},
  {Wehus}, {White}, {White}, {Wilkinson}, {Yvon}, {Zacchei}, \&
  {Zonca}}]{Pla16}
{Planck Collaboration}, {Ade}, P.~A.~R., {Aghanim}, N., {et~al.} 2016, \aap,
  594, A13, \dodoi{10.1051/0004-6361/201525830}

\bibitem[{{Rafiei-Ravandi} {et~al.}(2021){Rafiei-Ravandi}, {Smith}, {Li},
  {Masui}, {Josephy}, {Dobbs}, {Lang}, {Bhardwaj}, {Patel}, {Bandura},
  {Berger}, {Boyle}, {Brar}, {Cassanelli}, {Chawla}, {Dong}, {Fonseca},
  {Gaensler}, {Giri}, {Good}, {Halpern}, {Kaczmarek}, {Kaspi}, {Leung}, {Lin},
  {Mena-Parra}, {Meyers}, {Michilli}, {M{\"u}nchmeyer}, {Ng}, {Petroff},
  {Pleunis}, {Rahman}, {Sanghavi}, {Scholz}, {Shin}, {Stairs}, {Tendulkar},
  {Vanderlinde}, \& {Zwaniga}}]{rafi21}
{Rafiei-Ravandi}, M., {Smith}, K.~M., {Li}, D., {et~al.} 2021, arXiv e-prints,
  arXiv:2106.04354.
\newblock \doarXiv{2106.04354}

\bibitem[{{Ravi} {et~al.}(2016){Ravi}, {Shannon}, {Bailes}, {Bannister},
  {Bhandari}, {Bhat}, {Burke-Spolaor}, {Caleb}, {Flynn}, {Jameson}, {Johnston},
  {Keane}, {Kerr}, {Tiburzi}, {Tuntsov}, \& {Vedantham}}]{Rav16}
{Ravi}, V., {Shannon}, R.~M., {Bailes}, M., {et~al.} 2016, Science, 354, 1249,
  \dodoi{10.1126/science.aaf6807}

\bibitem[{{Reischke} {et~al.}(2021){Reischke}, {Hagstotz}, \&
  {Lilow}}]{Reisch21}
{Reischke}, R., {Hagstotz}, S., \& {Lilow}, R. 2021, \prd, 103, 023517,
  \dodoi{10.1103/PhysRevD.103.023517}

\bibitem[{{Saunders} {et~al.}(1991){Saunders}, {Frenk}, {Rowan-Robinson},
  {Efstathiou}, {Lawrence}, {Kaiser}, {Ellis}, {Crawford}, {Xia}, \&
  {Parry}}]{Saun91}
{Saunders}, W., {Frenk}, C., {Rowan-Robinson}, M., {et~al.} 1991, \nat, 349,
  32, \dodoi{10.1038/349032a0}

\bibitem[{{Schulz-Dubois} \& {Rehberg}(1981)}]{Schul81}
{Schulz-Dubois}, E.~O., \& {Rehberg}, I. 1981, Applied Physics, 24, 323,
  \dodoi{10.1007/BF00899730}

\bibitem[{{Shi} \& {Turner}(1998)}]{Shi98}
{Shi}, X., \& {Turner}, M.~S. 1998, \apj, 493, 519, \dodoi{10.1086/305169}

\bibitem[{{Shirasaki} {et~al.}(2017){Shirasaki}, {Kashiyama}, \&
  {Yoshida}}]{Shiras17}
{Shirasaki}, M., {Kashiyama}, K., \& {Yoshida}, N. 2017, \prd, 95, 083012,
  \dodoi{10.1103/PhysRevD.95.083012}

\bibitem[{{Sylos Labini}(2011)}]{Sylos11}
{Sylos Labini}, F. 2011, EPL (Europhysics Letters), 96, 59001,
  \dodoi{10.1209/0295-5075/96/59001}

\bibitem[{{Sylos Labini} {et~al.}(2009){Sylos Labini}, {Vasilyev}, \&
  {Baryshev}}]{Sylo09}
{Sylos Labini}, F., {Vasilyev}, N.~L., \& {Baryshev}, Y.~V. 2009, \aap, 496, 7,
  \dodoi{10.1051/0004-6361:200810575}

\bibitem[{{Takahashi} {et~al.}(2021){Takahashi}, {Ioka}, {Mori}, \&
  {Funahashi}}]{Takaha21}
{Takahashi}, R., {Ioka}, K., {Mori}, A., \& {Funahashi}, K. 2021, \mnras, 502,
  2615, \dodoi{10.1093/mnras/stab170}

\bibitem[{{The CHIME/FRB Collaboration} {et~al.}(2021){The CHIME/FRB
  Collaboration}, {:}, {Amiri}, {Andersen}, {Bandura}, {Berger}, {Bhardwaj},
  {Boyce}, {Boyle}, {Brar}, {Breitman}, {Cassanelli}, {Chawla}, {Chen},
  {Cliche}, {Cook}, {Cubranic}, {Curtin}, {Deng}, {Dobbs}, {Fengqiu}, {Dong},
  {Eadie}, {Fandino}, {Fonseca}, {Gaensler}, {Giri}, {Good}, {Halpern}, {Hill},
  {Hinshaw}, {Josephy}, {Kaczmarek}, {Kader}, {Kania}, {Kaspi}, {Landecker},
  {Lang}, {Leung}, {Li}, {Lin}, {Masui}, {Mckinven}, {Mena-Parra},
  {Merryfield}, {Meyers}, {Michilli}, {Milutinovic}, {Mirhosseini},
  {M{\"u}nchmeyer}, {Naidu}, {Newburgh}, {Ng}, {Patel}, {Pen}, {Petroff},
  {Pinsonneault-Marotte}, {Pleunis}, {Rafiei-Ravandi}, {Rahman}, {Ransom},
  {Renard}, {Sanghavi}, {Scholz}, {Shaw}, {Shin}, {Siegel}, {Sikora}, {Singh},
  {Smith}, {Stairs}, {Tan}, {Tendulkar}, {Vanderlinde}, {Wang}, {Wulf}, \&
  {Zwaniga}}]{chim21}
{The CHIME/FRB Collaboration}, {:}, {Amiri}, M., {et~al.} 2021, arXiv e-prints,
  arXiv:2106.04352.
\newblock \doarXiv{2106.04352}

\bibitem[{{Thornton} {et~al.}(2013){Thornton}, {Stappers}, {Bailes},
  {Barsdell}, {Bates}, {Bhat}, {Burgay}, {Burke-Spolaor}, {Champion}, {Coster},
  {D'Amico}, {Jameson}, {Johnston}, {Keith}, {Kramer}, {Levin}, {Milia}, {Ng},
  {Possenti}, \& {van Straten}}]{Tho13}
{Thornton}, D., {Stappers}, B., {Bailes}, M., {et~al.} 2013, Science, 341, 53,
  \dodoi{10.1126/science.1236789}

\bibitem[{{Walters} {et~al.}(2018){Walters}, {Weltman}, {Gaensler}, {Ma}, \&
  {Witzemann}}]{Walters18}
{Walters}, A., {Weltman}, A., {Gaensler}, B.~M., {Ma}, Y.-Z., \& {Witzemann},
  A. 2018, \apj, 856, 65, \dodoi{10.3847/1538-4357/aaaf6b}

\bibitem[{{Weinberg} {et~al.}(2004){Weinberg}, {Dav{\'e}}, {Katz}, \&
  {Hernquist}}]{Wein04}
{Weinberg}, D.~H., {Dav{\'e}}, R., {Katz}, N., \& {Hernquist}, L. 2004, \apj,
  601, 1, \dodoi{10.1086/380481}

\bibitem[{{Weinberg} {et~al.}(2013){Weinberg}, {Mortonson}, {Eisenstein},
  {Hirata}, {Riess}, \& {Rozo}}]{Wein13}
{Weinberg}, D.~H., {Mortonson}, M.~J., {Eisenstein}, D.~J., {et~al.} 2013,
  \physrep, 530, 87, \dodoi{10.1016/j.physrep.2013.05.001}

\bibitem[{{Xu}(2020)}]{Xu20}
{Xu}, S. 2020, \mnras, 492, 1044, \dodoi{10.1093/mnras/stz3092}

\bibitem[{{Xu} \& {Zhang}(2016{\natexlab{a}})}]{XZ16}
{Xu}, S., \& {Zhang}, B. 2016{\natexlab{a}}, \apj, 832, 199,
  \dodoi{10.3847/0004-637X/832/2/199}

\bibitem[{{Xu} \& {Zhang}(2016{\natexlab{b}})}]{XuZ16}
---. 2016{\natexlab{b}}, \apj, 824, 113, \dodoi{10.3847/0004-637X/824/2/113}

\bibitem[{{Xu} \& {Zhang}(2020{\natexlab{a}})}]{XuZfrb20}
---. 2020{\natexlab{a}}, \apjl, 898, L48, \dodoi{10.3847/2041-8213/aba760}

\bibitem[{{Xu} \& {Zhang}(2020{\natexlab{b}})}]{XZpul20}
---. 2020{\natexlab{b}}, \apj, 905, 159, \dodoi{10.3847/1538-4357/abc69f}

\bibitem[{{Yang} {et~al.}(2017){Yang}, {Luo}, {Li}, \& {Zhang}}]{Yang17}
{Yang}, Y.-P., {Luo}, R., {Li}, Z., \& {Zhang}, B. 2017, \apjl, 839, L25,
  \dodoi{10.3847/2041-8213/aa6c2e}

\bibitem[{{Yao} {et~al.}(2017){Yao}, {Manchester}, \& {Wang}}]{Yao17}
{Yao}, J.~M., {Manchester}, R.~N., \& {Wang}, N. 2017, \apj, 835, 29,
  \dodoi{10.3847/1538-4357/835/1/29}

\bibitem[{{Zehavi} {et~al.}(2002){Zehavi}, {Blanton}, {Frieman}, {Weinberg},
  {Mo}, {Strauss}, {Anderson}, {Annis}, {Bahcall}, {Bernardi}, {Briggs},
  {Brinkmann}, {Burles}, {Carey}, {Castander}, {Connolly}, {Csabai},
  {Dalcanton}, {Dodelson}, {Doi}, {Eisenstein}, {Evans}, {Finkbeiner},
  {Friedman}, {Fukugita}, {Gunn}, {Hennessy}, {Hindsley}, {Ivezi{\'c}}, {Kent},
  {Knapp}, {Kron}, {Kunszt}, {Lamb}, {Leger}, {Long}, {Loveday}, {Lupton},
  {McKay}, {Meiksin}, {Merrelli}, {Munn}, {Narayanan}, {Newcomb}, {Nichol},
  {Owen}, {Peoples}, {Pope}, {Rockosi}, {Schlegel}, {Schneider}, {Scoccimarro},
  {Sheth}, {Siegmund}, {Smee}, {Snir}, {Stebbins}, {Stoughton}, {SubbaRao},
  {Szalay}, {Szapudi}, {Tegmark}, {Tucker}, {Uomoto}, {Vanden Berk}, {Vogeley},
  {Waddell}, {Yanny}, \& {York}}]{Zehavi02}
{Zehavi}, I., {Blanton}, M.~R., {Frieman}, J.~A., {et~al.} 2002, \apj, 571,
  172, \dodoi{10.1086/339893}

\bibitem[{{Zhang}(2018)}]{Zha18}
{Zhang}, B. 2018, \apjl, 867, L21, \dodoi{10.3847/2041-8213/aae8e3}

\bibitem[{{Zheng} {et~al.}(2014){Zheng}, {Ofek}, {Kulkarni}, {Neill}, \&
  {Juric}}]{Zheng14}
{Zheng}, Z., {Ofek}, E.~O., {Kulkarni}, S.~R., {Neill}, J.~D., \& {Juric}, M.
  2014, \apj, 797, 71, \dodoi{10.1088/0004-637X/797/1/71}

\bibitem[{{Zhou} {et~al.}(2014){Zhou}, {Li}, {Wang}, {Fan}, \& {Wei}}]{Zhou14}
{Zhou}, B., {Li}, X., {Wang}, T., {Fan}, Y.-Z., \& {Wei}, D.-M. 2014, \prd, 89,
  107303, \dodoi{10.1103/PhysRevD.89.107303}

\bibitem[{{Zhu} {et~al.}(2018){Zhu}, {Feng}, \& {Zhang}}]{Zhu18}
{Zhu}, W., {Feng}, L.-L., \& {Zhang}, F. 2018, \apj, 865, 147,
  \dodoi{10.3847/1538-4357/aadbb0}

\end{thebibliography}

\end{document}